\documentclass{article}
\usepackage[utf8]{inputenc}
\usepackage[left=2cm,right=2cm,top=3.5cm,bottom=3.5cm]{geometry}
\usepackage{amsmath,amssymb,cases,color,authblk,url,enumerate,mathrsfs,arydshln,hyperref,empheq,graphicx,ulem,comment}

\newcommand{\beqn}{\begin{eqnarray}}
\newcommand{\eeqn}{\end{eqnarray}}
\newcommand{\barg}{\bar{g}}
\newcommand{\barphi}{{\bar{\phi}}}
\newcommand{\barX}{\bar{X}}
\newcommand{\bargamma}{\bar{\gamma}}
\newcommand{\barn}{\bar{n}}
\newcommand{\bara}{\bar{a}}
\newcommand{\barD}{\bar{\mathrm{D}}}
\newcommand{\barK}{\bar{K}}
\newcommand{\barR}{\bar{R}}
\newcommand{\barG}{\bar{G}}
\newcommand{\barT}{\bar{T}}
\newcommand{\bard}{\bar{D}}

\newcommand{\mrs}{{(\mathrm{S})}}
\newcommand{\mrd}{{\mathrm{d}}}
\newcommand{\V}{{(\mathrm{V})}}
\newcommand{\T}{{(\mathrm{T})}}
\newcommand{\bracketvarphi}{{(\varphi)}}
\newcommand{\bracketA}{{(A)}}
\newcommand{\bracketC}{{(C)}}
\newcommand{\lambdabar}{\lambda \kern -0.5em\raise 0.5ex \hbox{--}}
\newcommand{\barnabla}{\bar{\nabla}}
\newcommand{\Lie}{\pounds}
\newcommand{\nn}{\nonumber}
\newcommand{\4}{{(4)}}
\newcommand{\3}{{(3)}}

\makeatletter
\renewcommand{\@maketitle}{\newpage
\begin{flushright} \footnotesize YITP-22-92 \\ IPMU22-0043 \end{flushright}%report number
\null
\vskip 2em
\begin{center}
{\LARGE \@title \par} \vskip 1.5em {\large \lineskip .5em
\begin{tabular}[t]{c}\@author
\end{tabular}\par}
\vskip 1em {\large \@date} \end{center}
\par
\vskip 1.5em}
\makeatother
\title{Propagation of scalar and tensor gravitational waves in Horndeski theory}
\author[1]{Kei-ichiro Kubota}
\author[2]{Shun Arai}
\author[1,3]{Shinji Mukohyama}
\affil[1]{Center for Gravitational Physics, Yukawa Institute for Theoretical Physics, Kyoto University, 606-8502, Kyoto, Japan}
\affil[2]{Kobayashi-Maskawa Institute, Nagoya University, Nagoya 464-8602, Japan}
\affil[3]{Kavli Institute for the Physics and Mathematics of the Universe (WPI), The University of Tokyo Institutes for Advanced Study, The University of Tokyo, Kashiwa, Chiba, 277-8583, Japan} 
\date{\today}

\allowdisplaybreaks[1]

\begin{document}
\maketitle

\begin{abstract}
Gravitational waves travel through the distributions of matter and dark energy during propagation. For this reason, gravitational waves emitted from binary compact objects serve as a useful tool especially to probe the nature of dark energy. 
The geometrical optics approximation is a conventional way of investigating wave propagation. 
However, the approximation becomes less accurate as the wavelength approaches the curvature radius of the background, which can occur in generic situations.
In this paper, we suggest a formulation for higher-order corrections of the geometrical optics expansion, applied to Horndeski theory which accommodates many dark energy models. 
At the level of the background, assuming that the derivative of the scalar field is non-vanishing and timelike, we choose the time slices to coincide with the contours of the scalar field. This choice of the background time slices is advantageous as the sound cones of both scalar and tensor gravitational waves are upright with respect to the background time slices whenever the scalar field behaves as a perfect fluid. 
We then analyze the equations of motion for scalar and tensor components of gravitational waves at the leading and next-to-leading order in the geometrical optics expansion, deriving the evolution equations for their amplitudes under certain conditions. 
In particular, for Generalized Brans-Dicke theories, we find a simple description of equations for gravitational waves in terms of an effective metric.

\begin{description}
    \item[Keywords]gravitational wave, Horndeski theory, geometrical optics, dark energy.
\end{description}
\end{abstract}

\section{Introduction}
Gravitational waves (GWs), which were first detected by LIGO in 2015 \cite{LIGOScientific:2016aoc}, open a new way of studying the fundamental physics.
GWs generated by binary compact objects carry abundant physical information throughout the processes of emission and propagation.
The stage of emission provides opportunities for testing the theory of gravity in a strong gravitational field and nuclear physics in a high-density region, whereas the stage of propagation accumulates the effects of dark energy, i.e. the unidentified principal component of the present Universe at cosmological distances (see e.g.\cite{Ezquiaga:2018btd} for a review).
In this paper, we focus on the phenomenology of the propagation for the purpose of studying the nature of dark energy.
\par
Horndeski theory \cite{horndeski1974second,Deffayet:2013lga,Kobayashi:2011nu} is one of the covariant extensions of General Relativity (GR) by adding a single scalar field.
Horndeski theory is able to provide a number of interesting features on gravity and dark energy. 
Horndeski theory is the most general theory in which a Lagrangian density contains a metric and single scalar field and the covariant equations of motion (EoMs) are at most second-order differential equations in four dimensions. Horndeski theory contains gravity theories such as Jordan-Brans-Dicke theory \cite{PhysRev.124.925}, $f(R)$ gravity \cite{de2010f}, kinetic gravity branding \cite{Deffayet:2010qz,Pujolas:2011he} and major dark energy models; quintessence \cite{Ratra:1987rm,Caldwell:1997ii} and k-essence \cite{armendariz2001essentials,Chiba:1999ka}. Hence it is meaningful to study GW propagation in the framework of Horndeski theory in order to understand gravity and dark energy.
\par
Propagation of GWs in Horndeski theory has been studied in the literature. 
Most studies assume Minkowski spacetime e.g. \cite{Moretti:2019yhs} for $f(R)$ gravity or Friedmann-Lema\^{i}tre-Robertson-Waker (FLRW) spacetime as a background. %\cite{Moretti:2020kpp,Moretti:2019yhs,Akbarieh:2021vhv,Aslmarand:2021qwn,Akbarieh:2022ovn,Kazempour:2022let,Kazempour:2022giy}.
Considering the late-time Universe\footnote{There is an interesting phenomenon of Landau damping in a medium of non-collisional gas of particles \cite{Moretti:2020kpp}. This is not the case we study throughout the paper.}, GWs acquire two major aspects of dark energy in Horndeski theory that differ from the standard model of the Universe at late time i.e. GR with the cosmological constant. One is that the coupling with the dynamical dark-energy field changes the amplitude and phase of GWs (the original derivation is shown in \cite{Kobayashi:2011nu}. See \cite{Saltas:2014dha,Nishizawa:2017nef} for details of phenomenological formulation). The other is that an extra scalar polarization emerges other than the two standard tensor ones \cite{Hou:2017bqj}\footnote{It is known that static, spherically symmetric and asymptotically flat black holes cannot take non-trivial scalar field profiles in shift symmetric Horndeski theory without the coupling with Gauss-Bonnet invariant~\cite{Hui:2012qt,Sotiriou:2013qea,Creminelli:2020lxn}. However, the theorem is applicable to neither the propagation of GWs in galactic and intergalactic space nor the generation of GWs by binary black holes since the system is not static, spherically symmetric and asymptotically flat.}. Constraints have been placed on these features using observed GWs from binary compact objects (see e.g. \cite{Ezquiaga:2018btd} for a review). Currently, the constraint on the phase velocity with the binary neutron star merger GW170817 and the associating gamma ray burst GW170817A \cite{LIGOScientific:2017zic} is the most stringent\cite{Creminelli:2017sry,Ezquiaga:2017ekz,Sakstein:2017xjx,Baker:2017hug,Arai:2017hxj,Amendola:2017orw}. A constraint on the damping rate of GW amplitude has been obtained \cite{Arai:2017hxj} although it is still loose. The authors of \cite{Takeda:2021hgo} placed a constraint on the ratio of the scalar-mode amplitudes to the tensor-mode amplitudes for GWs with a mixture of the scalar and tensor polarizations.
Hence, it is expected that formulating the propagation of GWs for Horndeski theory provides a way to discriminate the models of dark energy from the standard model.
\par
In order to model a more precise universe in accord with observations, however, assuming the FLRW universe as a background spacetime is not necessarily adequate; it breaks at the scales where structures are formed. 
Although the measurements of cosmic microwave background and the observations of the large-scale structure have revealed that our Universe becomes spatially homogeneous and isotropic as it goes to larger distances \cite{Komatsu_2011,Planck:2015fie}, matter inhomogeneously distributes at galactic scales. Since GW sources are believed to be hosted in galaxies based on their population syntheses \cite{Stevenson:2015bqa,Mandel:2009nx,LIGOScientific:2021djp}, GWs indeed propagate through inhomogeneous spacetime, ending up with abandoning the FLRW spacetime. 
Hence, it is necessary to formulate a way to investigate the propagation of GWs including the inhomogeneities of matter and dark energy.
(See also \cite{Creminelli:2018xsv,Creminelli:2019nok,Creminelli:2019kjy,Gumrukcuoglu:2022bjy} for other implications of interactions between tensor GWs and scalar dark energy perturbations.)
\par
The geometrical optics approximation \cite{Isaacson:1968hbi,Isaacson:1968zza} (see e.g. \cite{maggiore2007gravitational}) is available to study the propagation of GWs on the background spacetime with inhomogeneities of matter and dark energy. 
A perturbative field can be separated into the amplitude and phase (called ``eikonal''), which varies rapidly. The approximation is consistent in any background spacetime whose variation scales are sufficiently longer than the wavelength of GWs.
Applying this approximation to GWs in GR, we can see that GWs propagate along the null geodesic, that the amplitude evolves to conserve the number of gravitons and that the polarization tensor is parallelly transported along the null geodesic.
In the case where the background spacetime has angular momentum, a rotation of polarization occurs as a result of the parallel transportation of the polarization.
This is known as gravitational Faraday rotation \cite{Dehnen:1973xa}.
Taking into account higher-order corrections to the approximation, we see that there is energy flow in a direction orthogonal to the propagation direction (see e.g. \cite{Anile:1976gq, Dolan:2018ydp} in the context of electromagnetic wave).
The approximation becomes less accurate as the wavelength is longer. GWs from heavier binary systems e.g. supermassive black hole binaries or ones emitted during the inspiral phase long before a merger may lie beyond the approximation. The case where a wavelength of a GW approaches the Schwarzschild radius of a lensing mass can no longer be treated with the geometrical optics due to non-negligible wave effects \cite{Takahashi:2003ix}. Towards this end, we suggest a formulation for the systematic computation of higher-orders of the geometrical optics expansion.
\par
The purpose of this paper is to suggest a formulation to systematically compute higher orders of the geometrical optics expansion in a subclass of Horndeski theory, studying the physical effects as much as possible toward higher orders in geometrical optics. At present, the solutions for the leading and next-to-leading order in the equations of motion for perturbations are known. Recently, the propagation of scalar and tensor modes in Horndeski theory has been studied in \cite{Garoffolo:2019mna,dalang2021scalar} (See also \cite{Reall:2021voz}) in the framework of geometrical optics approximation.
In \cite{Garoffolo:2019mna}, the authors consider the subclass of Horndeski theory in which speeds of sound for both scalar and tensor modes are the speed of light. 
In \cite{dalang2021scalar}, the authors derived the evolution equations for the amplitude of both scalar and tensor modes in the subclass of Horndeski theory in which the sound speed of tensor modes is the speed of light and the sound speed of the scalar field is arbitrary. In this paper, we shall study the latter subclass. 

\par
The rest of this paper is organized as follows.
In \S \ref{Horndeski_theory}, we present the theory that we study in this paper.
In \S \ref{geometric_optics_expansion}, we suggest the formulation to systematically compute higher orders of geometrical optics expansion taking into account inhomogeneities of matter and dark energy in our universe. 
In \S \ref{evo_eq_for_amplitude}, we study the leading and next-to-leading order terms of the equations for perturbative variables and derive the sound speeds and the evolution equations for the amplitudes.
Finally, we summarize and discuss our results in \S \ref{ch:summary}.
Throughout this paper, we use the following notation. The metric signature is $(-,+,+,+)$. The indices of tensors $a,b,c,d,\cdots$ run over 0 to 3, and the spatial indices $i,j,k,l,\cdots$ run over 1 to 3. $T_{(ab)}:=\frac{1}{2}(T_{ab}+T_{ba})$ denotes the symmetrization of the indices and $T_{[ab]}:=\frac{1}{2}(T_{ab}-T_{ba})$ denotes the anti-symmetrization of the indices. 
The numbers in the upper left of the Ricci tensor and Riemann tensor are the space/spacetime dimensions.
The left numbers in the upper right of the amplitude represent the order of perturbative expansion and the right numbers represent the order of geometrical optics expansion.
The numbers in the upper right of the other variables are the order of perturbative expansion.
The variables with a bar are the background variables.
We use the unit $c=1$.

\section{Horndeski theory with luminal tensor GW propagation}\label{Horndeski_theory}
We consider a certain subclass of Horndeski theory in which a tensor perturbation propagates at the speed of light.
The authors of \cite{Kobayashi:2011nu} derived the sound speeds of the scalar and tensor perturbations around FLRW background spacetime.
The sound speeds are written in Appendix \ref{app:Horndeski} for the whole Horndeski theory containing four free functions $G_{2,3,4,5}(\phi,X)$, where $\phi$ is the scalar field and $X\equiv -g^{ab}\partial_a\phi\partial_b\phi/2$. In the following, subscripts $\phi$ and $X$ denote the derivative with respect to $\phi$ and $X$, respectively. 
In general, $G_{4X} \neq 0$ or $G_{5} \neq 0$ are allowed, and then the propagation speed of the tensor mode around the FLRW spacetime is different from the speed of light.
\par
However, the simultaneous detection of GW170817 \cite{abbott2017gw170817} and GRB170817A \cite{goldstein2017ordinary,savchenko2017integral} has established that the tensor sound speed is the same as the speed of light with the allowed difference at the order of $10^{-15}$ \cite{abbott2017gw170817}.
Then models with $G_{4X} \neq 0$ or $G_{5} \neq 0$ have been severely constrained \cite{Creminelli:2017sry,Ezquiaga:2017ekz,Sakstein:2017xjx,Baker:2017hug,Arai:2017hxj}, otherwise it is necessary to invoke fine-tuning of the background dynamics (see e.g. \cite{Kase:2018aps,Copeland:2018yuh}).
We consider the subclasses of Horndeski theory that satisfy $G_{4X} = 0$ and $G_{5} = 0$ on any spacetime so that the tensor modes propagate at the speed of light without fine-tuning of the background dynamics. This subclass is given with the Lagrangian density as,
\beqn
 \mathcal{L}= \frac{1}{2\kappa^2}\bigl(G_{2}(\phi, X)+G_{3}(\phi, X) {\square} \phi+G_{4}(\phi) R \bigr) + \mathcal{L}^\mathrm{(matter)},
 \label{eq:LHorndeski}
\eeqn
where $\kappa^2 := {8\pi G_\mathrm{N}}$ and $G_\mathrm{N}$ is the Newton constant. We consider the matter component minimally-coupled to the metric. 
The variation of the action (\ref{eq:LHorndeski}) with respect to the metric gives the EoMs for the metric as
\beqn
  G_{ab} = \kappa^2 T_{ab}^\mathrm{(eff)} = \kappa^2\left(T_{ab}^\mathrm{(G_2)} + T_{ab}^\mathrm{(G_3)} + T_{ab}^\mathrm{(G_4)}+\frac{T_{ab}^\mathrm{(matter)}}{G_4}\right),
  \label{eq:Einsteineq}
\eeqn
where $G_{ab}$ is Einstein tensor and $T_{ab}^{(G_2)},T_{ab}^{(G_3)},T_{ab}^{(G_4)},T_{ab}^\text{(matter)}$ are 
\beqn
    T_{ab}^{(G_2)} &:=& \frac{1}{2 \kappa^2  G_4 } \Bigl[ G_{2X}\, \nabla_a \phi \nabla_b \phi+\, G_2\, g_{ab}\Bigr],\\
    T_{ab}^{(G_3)} &:=& \frac{1}{2 \kappa^2  G_4} \Bigl[ G_{3X}\, \Box \phi\, \nabla_a \phi \nabla_b \phi + \nabla_a G_3 \nabla_b \phi + \nabla_b G_3 \nabla_a \phi - g_{ab}\nabla_c G_3 \nabla^c \phi \Bigr],\\
    T_{ab}^{(G_4)} &:=& \frac{1}{\kappa^2 G_4} \Bigl[G_{4\phi}\left( \nabla_a \nabla_b \phi -  g_{ab}  \Box \phi \right) + {G_{4\phi\phi}}\left( \nabla_a \phi \nabla_b \phi + 2 X g_{ab} \right) \Bigr],\\
    T_{ab}^\text{(matter)}&:=&-\frac{2}{\sqrt{-\mathrm{det}(g_{cd})}}\frac{\delta S^\text{(matter)}}{\delta g^{ab}};\ \ S^\text{(matter)}:=\int \mathrm{d}^4 x \sqrt{-g}\mathcal{L}^\mathrm{(matter)}.
\eeqn
Variation of the action (\ref{eq:LHorndeski}) with respect to the scalar field gives the EoM for the scalar field as
\beqn
    \sum^4_{i=2}\mathcal{E}^{(G_i)}=\sum^4_{i=2}\Bigl[\mathcal{P}^{(G_i)} - \nabla^a \mathcal{J}^{(G_i)}_a\Bigr]=0,
    \label{eq:ScalarEoM}
\eeqn
where $\mathcal{E}^{(G_i)}:=\mathcal{P}^{(G_i)} - \nabla^a \mathcal{J}^{(G_i)}_a$, and
\beqn
    \mathcal{P}^{(G_2)}&:=&G_{2\phi},\\
    \mathcal{P}^{(G_3)}&:=&-(\nabla_a G_{3\phi})(\nabla^a \phi),\\
    \mathcal{P}^{(G_4)}&:=&G_{4\phi}R,\\
    \mathcal{J}^{(G_2)}_a&:=&-G_{2X}\nabla_a \phi,\\
    \mathcal{J}^{(G_3)}_a&:=&-G_{3X}\Box\phi\nabla_a \phi - G_{3X}\nabla_a X - 2 G_{3\phi}\nabla_a \phi,\\
    \mathcal{J}^{(G_4)}_a&:=&0.
\eeqn
If the Lagrangian density is shift-symmetric, i.e. if it is invariant under scalar field constant shift $\phi\to\phi+\mathrm{const.}$, then the EoM for $\phi$ is reduced to the conservation equation of the Noether current associated with the shift-symmetry, namely $\mathcal{J}_a^{(G_i)}$ becomes the contribution of $G_i$ to the Noether current and $\mathcal{P}^{(G_i)}$ identically vanishes (see, e.g., \cite{Kobayashi:2011nu}). 
In the following, we do not assume the shift symmetry and thus $\mathcal{P}^{(G_i)}$ is non-vanishing in general. 
In the following sections, we analyze these equations with geometrical optics expansion.

\section{Geometrical optics expansion with background unitary gauge}\label{geometric_optics_expansion}
We formulate a method to systematically compute higher orders of the geometrical optics expansion (see e.g. \cite{maggiore2007gravitational}) for the purpose of analyzing GWs in Horndeski theory. The geometrical optics expansion enables us to treat the propagation of GWs on the background with inhomogeneous distributions of matter and dark energy. To trace the time evolution of perturbations, it is useful to adopt a specific choice of time slicing, called the unitary gauge or the uniform scalar slicing, at the level of the background. This is because, with this choice of the background time slicing, the sound cones of both scalar and tensor gravitational waves are upright with respect to the time slicing whenever the scalar field behaves as a perfect fluid. For this reason, the behavior of perturbations can be understood intuitively. 
\par
We now briefly describe our methodology.
We split the metric and scalar field into a background and perturbations according to their spatial and temporal variation scales. We then assume that the derivative of the scalar field is non-vanishing and timelike at the level of the background. 
This allows us to choose the unitary gauge for the background. 
For perturbations, we employ the scalar-vector-tensor (SVT) decomposition based on the transformation properties under the spatial diffeomorphism on each constant background time surface. Then we divide the perturbative variables into the amplitudes and the phases for propagating modes.
\par
There are a couple of advantages of our method. The background unitary gauge (or the background uniform scalar slicing) makes the sound cones of both scalar and tensor gravitational waves upright with respect to the background time slicing whenever the scalar field behaves as a perfect fluid, while the sound speeds, i.e. the  opening angles of the sound cones, vary both spatially and temporally. Hence, the propagation of perturbations can be understood intuitively with this choice of the background time slicing. In particular, the phase evolution of the scalar GW is simple. On the other hand, the SVT decomposition helps to solve the dynamics of the GWs for individual propagating modes order by order in the geometrical optics expansion. Based on these procedures, we derive the equations of motion for the scalar and tensor GWs in each order of geometrical optics expansion. We shall show the details in the following subsections.

\subsection{Separation into background and perturbations}
We split the metric and scalar field into a background and perturbations as
\beqn
    g_{ab}=\barg_{ab}+h_{ab},\ \phi=\barphi+\varphi.
\eeqn
Here the bars denote the background variables, and $h_{ab}$ and $\varphi$ denote the perturbations. We assume that the norms of the perturbations are sufficiently smaller than those of the background, imposing
\beqn
    \|h_{ab}\|\ll\|\barg_{ab}\|,\ \|\varphi\|\ll\|\barphi\|.
\eeqn
We suppose that metric and scalar-field perturbations lie in the same order, and use the notation
\beqn
    h_{ab}\sim\varphi\sim\mathcal{O}(h),
\eeqn
where $h$ is a bookkeeping parameter that represents the smallness of the perturbations in comparison to the background variables $\bar{g}_{ab}\sim\bar{\phi}\sim\mathcal{O}(1)$\footnote{We adopt the normalization of the scalar field such that it is dimensionless.}.
Hereafter, the indices of the tensors defined on the background spacetime are raised and lowered by $\barg^{ab}$ and $\barg_{ab}$, where $\barg^{ab}$ is the inverse of $\barg_{ab}$.
\par
Let us introduce typical scales corresponding to the background and perturbations, respectively.
Suppose that the typical variation scales of $\barg_{ab}$ and $\barphi$ are controlled by $\mathcal{R}$,
\beqn
    \partial \barg_{ab} = \mathcal{O}(\mathcal{R}^{-1})\,, \quad \partial \barphi  = \mathcal{O}(\mathcal{R}^{-1})\,.
\eeqn
The perturbations $h_{ab}$ and $\varphi$ vary in scales of their wavelength. 
We thus suppose that their variations are controlled by the single scale $\lambda$, and we write $\lambda=2\pi \lambdabar$,
\beqn
    \partial h_{ab}\sim \frac{h_{ab}}{\lambdabar},\ \partial \varphi \sim \frac{\varphi}{\lambdabar}.
\eeqn
We assume that the wavelength $\lambda$ is sufficiently smaller than the typical variation scale $\mathcal{R}$ of the background \footnote{When this assumption does not hold, geometrical optics approximation is invalid. In such a situation, we need to use wave optics (see e.g. \cite{Takahashi:2003ix})} for validity of the geometrical optics expansion. In order to make this statement quantitative we introduce
\beqn
    \epsilon:=\frac{\lambdabar}{\mathcal{R}}\ll 1, \label{epsilon<<1}
\eeqn
as a bookkeeping parameter. 
Let us take an example to demonstrate that our assumptions are reasonable and applicable to the observed Universe and the first gravitational wave event GW150914.
We estimate the radius of curvature generated by our galaxy.
Supposing that deviation from GR is not too significant, the Einstein equations give the estimate of the radius of curvature generated by the density $\rho$ as $\mathcal{R}\sim \sqrt{c^2/(G_\mathrm{N}\rho)}$. Substituting the average density of our galaxy $\rho \sim 10^{-22}\ \mathrm{kg\ m^{-3}}$ into this expression of $\mathcal{R}$, the radius of curvature generated by the matter density is estimated to be $\mathcal{R} \sim 10^{21}\ \mathrm{m}$.
On the other hand, the observed frequency of GW150914 is $f \sim 10\ \mathrm{Hz}$ and thus the wavelength is $\lambda \sim 10^{7}\ \mathrm{m}$ while the strain is $h \sim 10^{-21}$\cite{LIGOScientific:2016aoc}. Therefore, our assumptions are valid for GW150914 when inhomogeneities at the galactic scales are included in the background. The frequencies of the GWs detected by LIGO and Virgo are in the range of Hz to kHz and the amplitude is at most $10^{-21}$. Thus, we conclude that our assumptions are valid in many situations. 
\par
Furthermore, we assume that the sound speeds of the perturbations are $\mathcal{O}(1)$ so that the period and wavelength lie in the same order. This assumption combined with (\ref{epsilon<<1}) implies 
\beqn
    \frac{\text{(period of perturbations)}}{\text{(time variation scale of background variables)}} \ll 1\,.
\eeqn
\par
We introduced the two small bookkeeping parameters. One is the smallness parameter $h$ for the perturbative expansion, and the other is the shortness parameter $\epsilon$ for the geometrical optics expansion. In general, the hierarchy between $h$ and $\varepsilon$ is not a priori determined. To proceed further we assume the following hierarchy between $h$ and $\epsilon$,
\beqn
    h \ll \epsilon \ll 1.
    \label{eq:hierarchy}
\eeqn
Under this assumption (\ref{eq:hierarchy}), the back reaction of the GWs to the background metric can be negligible at least up to the next-to-leading order considered in this paper. Similarly to the above demonstration for GW150914, we confirm that this assumption is also valid for typical GWs detected by LIGO and Virgo when inhomogeneities at galactic scales are included in the background.
Considering Horndeski theory, the hierarchy \eqref{eq:hierarchy} allows the arbitrary functions $G_{2,3,4,5}$ to be expanded in $\delta X$ given by 
\beqn
    \delta X := X - \barX = \frac{1}{2}(h^{ab}\barnabla_a\barphi\barnabla_b\barphi)-\barnabla^a \varphi \barnabla_a \barphi + \cdots = \mathcal{O}(h \epsilon^{-1}) \ll 1.
\eeqn
For the subclass of Horndeski theory we consider in this paper, the expansion with respect to $\delta X$ is needed for $G_2$ and $G_3$. As already stated earlier, the quantities with the bar represent the background value.

\subsection{Foliation of background spacetime by scalar field}

\begin{figure}[h]
  \centering
  \includegraphics[width=15cm]{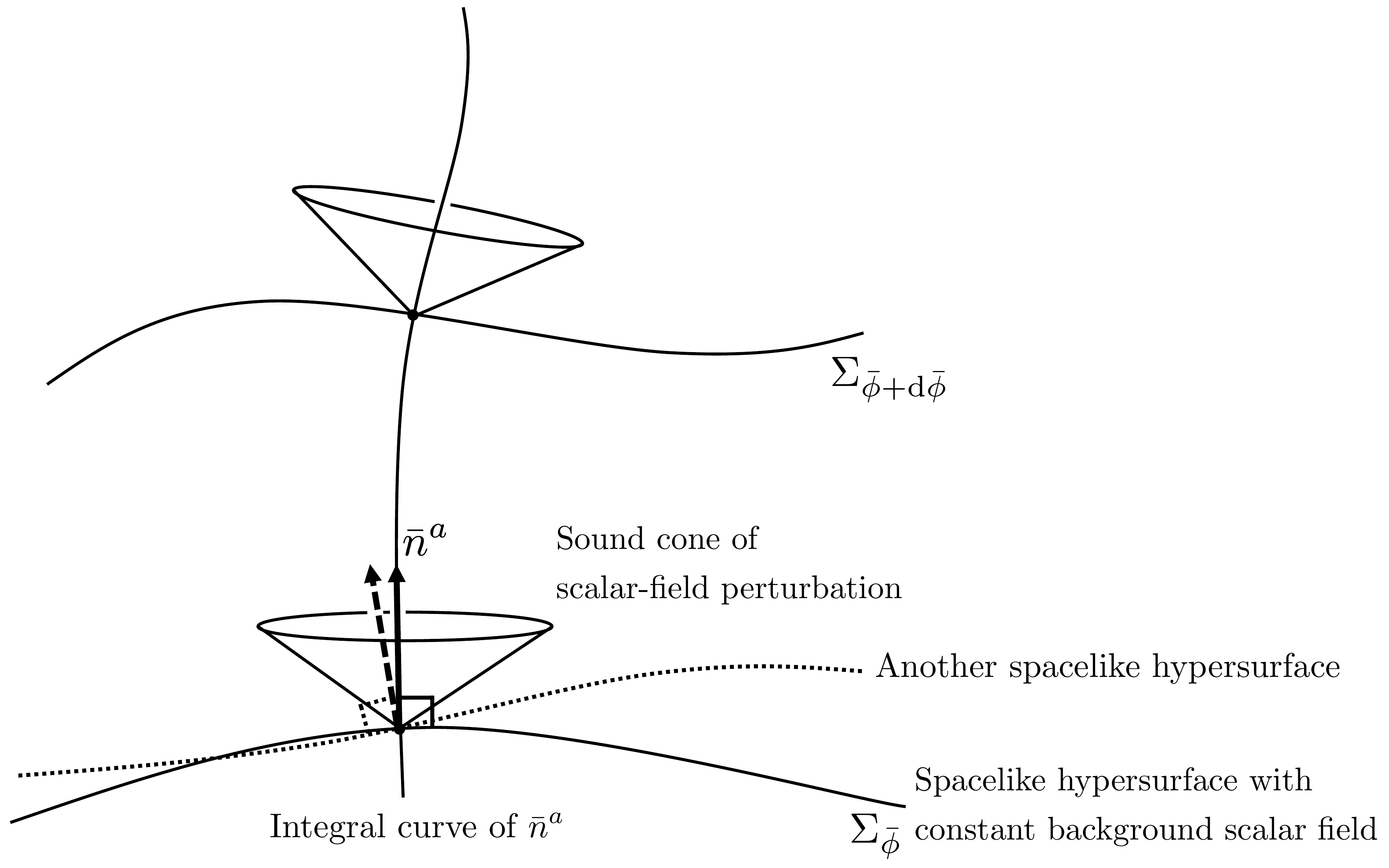}
  \caption{By choosing the time coordinate to set $\barphi=\barphi(t)$, we foliate the background spacetime by spacelike hypersurfaces with normal vectors having the same direction as the axis of the sound cone whenever the scalar field behaves as a perfect fluid. The sound cone is tilted with respect to the normal vector of other hypersurfaces even when the scalar field behaves as a perfect fluid.}
  \label{fg:ADMandSoundCone}
\end{figure}

Recall that the derivative of the scalar field is non-vanishing and timelike at the level of the background as we assumed. We can therefore choose the background time coordinate so that the background scalar field is a function of the time only and independent of spatial coordinates \footnote{Hence at the level of the background, the background scalar field is homogeneous on each constant time hypersurface. Note that geometry and matter are not necessarily homogeneous on each constant time hypersurface.}. 
An advantage of this choice is that, as we shall see later, the sound cones for both scalar and tensor gravitational waves are always upright, i.e. the axis of each sound cone is normal to the constant time hypersurface (see Fig. \ref{fg:ADMandSoundCone}), whenever the scalar field behaves as a perfect fluid. 
The unit vector normal to each constant time hypersurface is 
\beqn
    \barn^a = \frac{\barnabla^a \barphi}{\sqrt{2\bar{X}}}.
\eeqn
The background induced metric on the hypersurface $\Sigma_{\barphi}$ is then 
\beqn
    \bargamma_{ab}:=\barg_{ab} + \barn_a \barn_b.
\eeqn
By definition the background induced metric is of rank-$3$ and the normal vector $\barn^a$ is in its kernel, i.e. $\bargamma_{ab}\barn^a=0$. Hence, the background induced metric acts as a projection operator, i.e. $\bargamma^{a}_{\ b}=\barg^{ac} \bargamma_{cb}=\bargamma^{a}_{\ c} \bargamma^{c}_{\ b}$. 
\par
Using the acceleration $\bara^a:=\barn^c \barnabla_c \barn^a $ and the extrinsic curvature $\barK_{ab}:=\Lie_{\barn}\bargamma_{ab}/2=\bargamma_{a}^{\ c} \barnabla_c \barn_b$, the covariant derivative of the normal vector $n^a$ can be written as
\beqn
    \barnabla_a \barn_b=\barK_{ab}-\barn_a \bara_b.
    \label{eq:GradNormal1}
\eeqn
\par
To define the Riemann tensor of the hypersurface $\Sigma_{\barphi}$, we introduce the covariant derivative operator $\bar{\mathrm{D}}_a$ associated with the induced metric $\bargamma_{ab}$ which maps spatial tensors to spatial tensors.
The covariant derivative operator $\bar{\mathrm{D}}_a$ can be defined to act on any spatial tensor $S^{b_1 b_2 \cdots b_k}_{\ \ \ \ c_1 c_2  \cdots c_l}$ as
\beqn
    \bar{\mathrm{D}}_a S^{b_1 b_2 \cdots b_k}_{\ \ \ \ c_1 c_2  \cdots c_l} := \bargamma^{\ c}_{a}\bargamma_{c_1}^{\ e_1}\bargamma_{c_2}^{\ e_2}\cdots\bargamma_{c_l}^{\ e_l}\bargamma_{\ d_1}^{b_1}\bargamma_{\ d_2}^{b_2}\cdots\bargamma_{\ d_k}^{b_k}\barnabla_c S^{d_1 d_2 \cdots d_k}_{\ \ \ \ e_1 e_2  \cdots e_l}.
    \label{eq:defbarD}
\eeqn
The covariant derivative operator $\barD_a$ satisfies the metric compatibility of the induced metric, that is, $\barD_c \bargamma_{ab}=0$. 
The 3-dimensional Riemann tensor of the hypersurface $\Sigma_{\barphi}$,  ${^{(3)}}\bar{R}_{abcd}$ can be defined as a variable that stands for the noncommutativity of the covariant derivative operator $\barD_a$:
\beqn
    {^{(3)}}\barR_{abc}^{\ \ \ \ d}S_d:=(\barD_a \barD_b - \barD_b \barD_a)S_c,
\eeqn
where $S_a$ denotes a spatial vector. Using Gauss equation, Codazzi equation, and  Ricci equation (see e.g. \cite{Gourgoulhon:2007ue}), the 4-dimensional Riemann tensor ${^{(4)}}\bar{R}_{abcd}$ is written by the acceleration, extrinsic curvature, and its Lie derivative along $\barn^a$ as
\beqn
    {^\4}\bar{R}_{abcd}
    &=&{^\3}\bar{R}_{abcd}+2\bar{K}_{a[c} \bar{K}_{d]b}-4\left(\barD_{[a}\bar{K}_{b][c}\right)\bar{n}_{d]}-4\left(\barD_{[c} \bar{K}_{d][a}\right) \barn_{b]}\nonumber \\
    &&+4 \barn_{[a} \left( {\Lie}_{\barn} \bar{K}_{b][c} \right) \barn_{d]}-4 \barn_{[a} \barK_{b]}^{\ e} \barK_{e[c} \barn_{d]}-4 \barn_{[a} \bara_{b]} \bara_{[c} \barn_{d]}-4 \barn_{[a}\left( \barD_{b]} \bara_{[c} \right) \barn_{d]}.
    \label{eq:GaussCodazziRicci}
\eeqn

\subsection{Decomposition of metric perturbations}
We decompose the metric perturbation in the way of the SVT decomposition following \cite{Pitrou:2013hga} in order to simplify our analysis. We denote the time-time component of the metric perturbation as 
\beqn
    A:=h_{ab}\barn^a \barn^b,
\eeqn
the time-space component of the metric perturbation as 
\beqn
    B_a:=-h_{bc}\barn^b \bargamma^c_{\ a},
\eeqn
the trace part of the space-space component of the metric perturbation as 
\beqn
    C:=\frac{1}{3}h_{cd}\bargamma^c_{\ a}\bargamma^d_{\ b}\bargamma^{ab}=\frac{1}{3}h_{cd}\bargamma^{cd},
\eeqn
and the traceless part of the space-space component of the metric perturbation as
\beqn
    F_{ab}:=h_{ef}\bargamma^e_{\ c}\bargamma^f_{\ d}\left(\bargamma^c_{\ a}\bargamma^d_{\ b}-\frac{1}{3}\bargamma^{cd}\bargamma_{ab}\right)=h_{ef}\left(\bargamma^e_{\ a}\bargamma^f_{\ b}-\frac{1}{3}\bargamma^{ef}\bargamma_{ab}\right).
\eeqn
By using these variables, the metric perturbation can be expressed as
\beqn
    h_{ab}=A\barn_a \barn_b + B_a \barn_b + B_b \barn_a + C\bargamma_{ab} + F_{ab}.
\eeqn
Furthermore, $B_a$ is decomposed into the scalar mode $B$ and vector mode $B_a^\V$ as
\beqn
    B_a=\barD_a B + B^\V_a,
\eeqn
where $B_a^\V$ is a transverse component, i.e., $\barD^a B^\V_a=0$, and $F_{ab}$ is decomposed into the scalar mode $E$, vector mode $E_a^\V$, and tensor mode $E^\T_{ab}$ as
\beqn
    F_{ab}&=& \left( \bard_a \bard_b - \frac{1}{3}\bargamma_{ab} \bargamma^{cd}\bard_c \bard_d \right)E + \bard_a E^{(\mathrm{V})}_b + \bard_b E^{(\mathrm{V})}_a + E_{ab}^\T,
\eeqn
where $E_a^\V$ and $E^\T_{ab}$ are respectively transverse and transverse-traceless components, i.e., $\barD^a E^\V_a = 0$ and $\barD^a E^\T_{ab} = 0 = \bargamma^{ab}E^\T_{ab}$.
\par
We comment that the decomposition is efficient to proceed the geometrical optics expansion. As we shall see in \S 4, at the leading order in the geometrical optics expansion all covariant derivatives in the field equations for the perturbations act on the phases of the perturbation variables, and hence in particular the spatial vector or spatial tensor components produced by the derivative acted on the background variables do not show up. Thus, at the leading order in the geometrical optics expansion the EoMs for the scalar perturbations and the tensor perturbation defined by the SVT decomposition are decoupled from each other, and we obtain the individual sound speed.

\subsubsection{Gauge fixing}
We consider the infinitesimal gauge transformation $x^a \to x'^{a} = x^a + \xi^a$. We also decompose the gauge parameter $\xi^a$ into scalar and vector components as 
\beqn
    \xi^a = T\barn^a + \barD^a L + L^{\V a},
\eeqn
where $\barD^a L^\V_a =0$. Under the infinitesimal gauge transformation, the metric perturbation transforms as
\beqn
    h_{ab}\to h'_{ab}=h_{ab}-\Lie_\xi \barg_{ab}.
\eeqn
The transformation of each component of the metric is written in Appendix \ref{gaugetrnsformations}. 
\par
Since we discuss the perturbations around the arbitrary background and the spatial metric cannot be separated into a part depending on time coordinates and a part depending on spatial coordinates, the commutator of time derivative and spatial derivative cannot generally be zero, i.e. $[\partial_t, \barD_i] \neq 0$. As a result, if we take the unitary gauge $\varphi=0$, then the EoMs that complicatedly include the scalar modes obtained from the perturbed metric should need to derive the EoM of the scalar propagating mode. On the other hand, if we do not take the unitary gauge then we can extract the EoM for the scalar mode directly from the perturbed scalar field EoM, which already shows decoupling between the scalar and tensor modes at the leading order in the geometric optics expansion. Thus, we do not take the unitary gauge. 
\par
In this paper, we take a gauge in which the following components are eliminated using gauge degrees of freedom,
\beqn
   B = E = 0,\  E^{\V}_a = 0.
   \label{eq:gauge_condition}
\eeqn
Under this gauge fixing, the metric perturbation can be written as
\beqn
    h_{ab}= A \barn_a \barn_b + B^\V_a \barn_b  + B^\V_b \barn_a + \barn_b B^\V_a + C\bargamma_{ab} + E^\T_{ab}.
    \label{eq:SVT}
\eeqn
When we perturbatively expand the metric $h_{ab} = \sum_{i=1}^\infty h^{(i)}_{ab} h^i$, we can eliminate $B^{(i)}, E^{(i)}, E^{(i)}_a$ from the $i$th order metric perturbation $h^\text{(i)}_{ab}$ using the $i$-th order gauge parameter $\xi^{(i)a}$\cite{Brizuela:2008ra}. Thus the $i$th order metric perturbation can be written in the form 
\beqn
    h^{(i)}_{ab}= A^{(i)} \barn_a \barn_b  +  B^{\V(i)}_a \barn_b + B^{\V(i)}_b \barn_a + C^{(i)}\bargamma_{ab} + E^{\T(i)}_{ab}
\eeqn
Hereafter, we take the gauge in which all orders of $B, E, E^{\V}_a$ are zero.

\subsubsection{Amplitudes and phases for individual modes}
We divide the perturbations into the part $\mathcal{A}^\text{(m)}\exp(i\theta^\T/\epsilon)$ propagating at the speed of light and the part $\mathcal{B}^\text{(m)}\exp(i\theta^\mrs/\epsilon)$ propagating at the sound speed of the scalar field, where $\text{m}=\{\varphi,A,C,\mathrm{V},\mathrm{T}\}$. We write the type of perturbations on the upper right,
\beqn
    \varphi &=& \mathcal{A}^\bracketvarphi \exp\left(i\theta^\T/\epsilon \right) + \mathcal{B}^\bracketvarphi \exp\left(i\theta^\mrs/\epsilon \right),\\
    A &=& \mathcal{A}^\bracketA \exp\left(i\theta^\T/\epsilon \right) +\mathcal{B}^\bracketA \exp\left(i\theta^\mrs/\epsilon \right),\\
    C &=& \mathcal{A}^\bracketC \exp\left(i\theta^\T/\epsilon \right) + \mathcal{B}^\bracketC \exp\left(i\theta^\mrs/\epsilon \right),\\
    B^\V_{a} &=& \mathcal{A}^\V_a \exp\left(i\theta^\T/\epsilon \right) + \mathcal{B}^\V_a \exp\left(i\theta^\mrs/\epsilon \right) ,\\
    E^\T_{ab} &=& \mathcal{A}^\T_{ab} \exp\left(i\theta^\T/\epsilon \right) + \mathcal{B}^{\T}_{ab} \exp\left(i\theta^\mrs/\epsilon \right),
\eeqn
where $\mathcal{A}^\text{(m)}$ and $\mathcal{B}^\text{(m)}$ denote the amplitudes of each part.
We expand the amplitude $\mathcal{A}^\text{(m)}$ and $\mathcal{B}^\text{(m)}$ with respect to the bookkeeping parameter $h$ for the perturbative expansion and the bookkeeping parameter $\epsilon$ for the geometrical optics expansion,
\beqn
    \mathcal{A}^\text{(m)}=\sum_{i=1}^\infty \sum_{j=0}^\infty \mathcal{A}^{\text{(m)}(i,j)} h^i \epsilon^j ,\ \mathcal{B}^\text{(m)}=\sum_{i=1}^\infty \sum_{j=0}^\infty \mathcal{B}^{\text{(m)}(i,j)} h^i \epsilon^j .
\eeqn

\subsection{Equations of motion}
We divide the EoMs into a low-frequency part and a high-frequency part. For this purpose we introduce a scale $\ell$, which is larger than the wavelength $\lambda$ and shorter than the typical variation scale $\mathcal{R}$ of the background, i.e.
\beqn
    \lambdabar \ll \ell \ll \mathcal{R}.
\eeqn
\par
Averaging the EoMs over the scale $\ell$ (e.g. Brill-Hartle average \cite{brill1964method}) yields the equations of motion that determine the behavior of the background variables \cite{Isaacson:1968hbi,Isaacson:1968zza}. By averaging the EoMs (\ref{eq:Einsteineq}) over $\ell$, we obtain
\beqn
    0=\Bigl<G_{ab}-\kappa^2 T^{\mathrm{(eff)}}_{ab}\Bigr>_\ell =\barG_{ab}-\kappa^2 \barT^{\mathrm{(eff)}}+ \mathcal{O}(h^2 \epsilon^{-2}).
    \label{eq:largescaleScalarEoM}
\eeqn
By averaging the EoM for the scalar field (\ref{eq:ScalarEoM}) over $\ell$, we obtain
\beqn
    0=\Bigl<{\mathcal{E}}^{(G_2)}+{\mathcal{E}}^{(G_3)}+{\mathcal{E}}^{(G_4)}\Bigr>_\ell=\bar{\mathcal{E}}^{(G_2)}+\bar{\mathcal{E}}^{(G_3)} + \bar{\mathcal{E}}^{(G_4)} + \mathcal{O}(h^2 \epsilon^{-2}).
\eeqn
\par
The equations after subtracting the part averaged over $\ell$ from the original equations become the equations for the perturbations. Subtracting the averaged part, the original EoMs (\ref{eq:Einsteineq}) become
\beqn
    0=G_{ab}-\kappa^2 T^{\mathrm{(eff)}}_{ab}-\Bigl<G_{ab}-\kappa^2 T^{\mathrm{(eff)}}_{ab}\Bigr>_\ell = \Bigl[G_{ab}-\kappa^2 T^{\mathrm{(eff)}}_{ab}\Bigr]^{(1)}+\mathcal{O}(h^2 \epsilon^{-2}),
    \label{eq:smallscaleEinstein}
\eeqn
where $[\cdots]^{(1)}$ denotes the first order in perturbative expansion. The original EoM for the scalar field (\ref{eq:ScalarEoM}) becomes
\beqn
    0={\mathcal{E}}^{(G_2)}+{\mathcal{E}}^{(G_3)}+{\mathcal{E}}^{(G_4)}-\Bigl<{\mathcal{E}}^{(G_2)}+{\mathcal{E}}^{(G_3)}+{\mathcal{E}}^{(G_4)}\Bigr>_\ell = \Bigl[{\mathcal{E}}^{(G_2)}+{\mathcal{E}}^{(G_3)}+{\mathcal{E}}^{(G_4)}\Bigr]^{(1)}+\mathcal{O}(h^2 \epsilon^{-2}).
    \label{eq:smallscaleScalarEoM}
\eeqn
In this paper, we do not take into account matter fluctuations on scales comparable to or smaller than the wavelength of the gravitational waves. Thus, the matter energy-momentum tensor does not have the high-frequency part, i.e.
\beqn
    T^\text{(matter)}_{ab}-\Bigl< T^\text{(matter)}_{ab} \Bigr>_\ell = 0.
\eeqn
\par
To study the individual behavior of each mode, i.e. the sound speed and evolution of amplitude in the direction of the wave vector, the high-frequency part of the EoMs (\ref{eq:smallscaleEinstein}) have to be decomposed.
We project the perturbed EoMs for the metric into time-time, time-space and space-space components in the manner of the SVT decomposition.
Projecting Eq. (\ref{eq:smallscaleEinstein}) by $\barn^a \barn^b,\ \barn^a \bargamma^{b}_{\ c},\ \bargamma^{ab}$, and $ (\bargamma^a_{\ c}\bargamma^b_d-\bargamma_{cd}\bargamma^{ab}/3)$, it is decomposed into the perturbative equations of the Hamiltonian constraint, the momentum constraint, and the trace and traceless part of the evolution equation for the extrinsic curvature, respectively.
\par
For example, let us perform the perturbative expansion of the Hamiltonian constraint as
\beqn
    0 &=& \Bigl[\barG_{ab}-\kappa^2 \barT^{\text{(eff)}}_{ab}\Bigr]\barn^a \barn^b 
    +\Bigl[G_{ab}-\kappa^2 T^{\text{(eff)}}_{ab}\Bigr]^{(1)}\barn^a \barn^b 
    +\Bigl[\barG_{ab}-\kappa^2 \barT^{\text{(eff)}}_{ab}\Bigr]n^{(1)}{}^a n^{(1)}{}^b + \mathcal{O}(h^2\epsilon^{-2}),
\eeqn
where $n^{(1)a}$ denotes the first order part of $n^a=\nabla^a\phi/\sqrt{2X}$ in the perturbative expansion.
Since the first and third terms of this equation are zero up to $\mathcal{O}(h^2\epsilon^{-2})$ by using Eq. (\ref{eq:largescaleScalarEoM}), this equation yields
\beqn
    0=\Bigl[G_{ab}-\kappa^2 T^{\text{(eff)}}_{ab}\Bigr]^{(1)}\barn^a \barn^b + \mathcal{O}(h^2\epsilon^{-2}).
\eeqn
Thus Eq. (\ref{eq:smallscaleEinstein}) projected by $\barn^a \barn^b$ is the high-frequency part of the Hamiltonian constraint equation. 
Similarly, the high-frequency part of the momentum constraint and the trace and traceless part of the evolution equation for the extrinsic curvature respectively are derived by projecting Eq. (\ref{eq:smallscaleEinstein}) by $\barn^a \bargamma^{b}_{\ c},\ \bargamma^{ab}$, and $ (\bargamma^a_{\ c}\bargamma^b_d-\bargamma_{cd}\bargamma^{ab}/3)$.
The linearized Einstein tensor is presented in Appendix \ref{app:LEinsteinTensor}.
\par
We mention the physical meaning for each order of the EoMs. As we shall explicitly show in the next section, he leading order $\mathcal{O}(h^1 \epsilon^{-2})$ terms of the high-frequency part turns out to be the equation for the wave vector. The next-to-leading order $\mathcal{O}(h^1 \epsilon^{-1})$ terms of the high-frequency part turns out to be the evolution equation for the leading amplitudes $\mathcal{A}^{(1,0)}$, $\mathcal{B}^{(1,0)}$. Let us argue the next-to-next-to-leading order. The terms in such order of the high frequency part turns out to be the evolution equation for the next-to-leading amplitude $\mathcal{A}^{(1,1)}$, $\mathcal{B}^{(1,1)}$. 
The candidates for the next-to-next-to-leading order are $\mathcal{O}(h^1\epsilon^0)$ and $\mathcal{O}(h^2\epsilon^{-2})$, and we need more information about the hierarchy between $h$ and $\epsilon$ to know to identify the next-to-next-to-leading order. If $\epsilon^2 \ll h$, then the next-to-next-to-leading order is $\mathcal{O}(h^2\epsilon^{-2})$, if $\epsilon^2 \gg h$, then the next-to-next-to-leading order is $\mathcal{O}(h^1\epsilon^{0})$. Similarly, assuming magnitude relations between $h$ and $\epsilon^2$ allows us to find the evolution of further higher order amplitudes expressed in terms of lower order amplitudes. In this way, we can systematically obtain the equations for the higher order amplitudes.

\section{Scalar and tensor gravitational waves in geometrical optics}\label{evo_eq_for_amplitude}
We expand the EoMs in the previous section by $h$ and $\epsilon$ under the assumption Eq.~\eqref{eq:hierarchy}. 
If the assumption (\ref{eq:hierarchy}) holds, the leading order is $\mathcal{O}(h\epsilon^{-2})$, and the next-to-leading order is $\mathcal{O}(h\epsilon^{-1})$. We shall obtain the conditions for the wave vectors from the leading order and the evolution of the amplitudes of the variables $\varphi$ and $E^{(T)}_{ab}$ in the direction of the wave vector from the next-to-leading order, respectively. In addition, we shall find the relations among the variables $A$, $C$, and $B^{(V)}_a$ to satisfy all the EoMs consistently.
\par
For analyzing the subclass of Horndeski theory given in Sec.~\ref{Horndeski_theory}, we consider the two cases further specifying $G_{3}(\phi,X)$ and $G_4(\phi)$ on the background. In the case with $\barG_{3X} \neq 0$ or $\barG_{4\phi} \neq 0$, the scalar modes of the metric and scalar field lie in the same order, otherwise they are not. Therefore, the following two cases,
\begin{align}
    \text{Case with $h^\text{(scalar)}_{ab}\sim\epsilon\varphi$:}&\ \ \mathcal{L}= \frac{1}{2\kappa^2}\bigl( G_{2}(\phi, X) + R \bigr) + \mathcal{L}^\mathrm{(matter)}, \label{eq:LagrangianCaseI}\\
    \text{Case with $h^\text{(scalar)}_{ab}\sim\varphi$:}&\ \ \mathcal{L}= \frac{1}{2\kappa^2}\bigl(G_{2}(\phi, X)+G_{3}(\phi, X) {\square} \phi+G_{4}(\phi) R \bigr) + \mathcal{L}^\mathrm{(matter)};\ \barG_{3X} \neq 0 \text{ or/and } \barG_{4\phi} \neq 0, \label{eq:LagrangianCaseII}
\end{align}
should be separately considered.

Let us comment on the case with $G_3 = G_3(\phi)$ and $G_{4\phi}=0$. This case belongs to the first case and is equivalent to the case with $G_3=0$ and $G_{4\phi}=0$ for the following reason. The $G_{3}$ term in the action in this case is written as
\beqn
    \int \sqrt{-g} \mrd^4 x \Bigl(G_{3}(\phi)\Box \phi \Bigr)=\int \sqrt{-g} \mrd^4 x \Bigl(-(\nabla_a G_{3}(\phi))\nabla^a \phi \Bigr)=\int \sqrt{-g} \mrd^4 x \Bigl(G_{3\phi}(\phi)2X \Bigr),
\eeqn
up to a boundary term, which does not contribute to the bulk EoMs. Redefining $G_{2}(\phi,X)$ as $G'_{2}(\phi, X):=G_{2}(\phi,X)+2XG_{3\phi}(\phi)$, $G_{3}(\phi)$ is absorbed by $G_{2}(\phi,X)$. 

\subsection{Case with $h^\mathrm{(scalar)}_{ab}\sim\epsilon\varphi$: k-essence}\label{sec:caseI}
We investigate the leading and next-to-leading order of Eq. (\ref{eq:smallscaleEinstein}) and (\ref{eq:smallscaleScalarEoM}) for the model in which the Lagrangian density is Eq. (\ref{eq:LagrangianCaseI}). In this case, as we shall see below, the scalar modes of the metric perturbations are not in the same order as the scalar field perturbation.

\subsubsection{Leading order}
First, we investigate the leading order terms of Eq. (\ref{eq:smallscaleEinstein}) and (\ref{eq:smallscaleScalarEoM}). We reduce the number of derivatives acting on the perturbed variables as much as possible using the transverse conditions $\barD^a B^\V_a = 0$ and $\barD^a E^\T_{ab} = 0$, and use Eq. (\ref{eq:GradNormal1}), (\ref{eq:GaussCodazziRicci}). Then each mode decouples because the leading order terms do not contain the background components with derivatives such as the acceleration and extrinsic curvature.
\par
The linearized energy-momentum tensor does not contain the covariant derivative acting twice on the perturbations. Thus the leading order terms of the time-time components of the metric EoMs and the trace part of the space-space components yield trivial conditions on the scalar-mode amplitudes as
\beqn
    \mathcal{A}^{\bracketA (1,0)}=\mathcal{A}^{\bracketC (1,0)}=\mathcal{B}^{\bracketC (1,0)}=\mathcal{B}^{\bracketA (1,0)}=0,
\eeqn
where $\bargamma^{ab}\barnabla_a \theta^\mrs \barnabla_b \theta^\mrs \neq 0$, $\bargamma^{ab}\barnabla_a \theta^\T \barnabla_b \theta^\T \neq 0$ because we consider propagating modes. This means that the scalar modes contained in the metric are smaller than the amplitudes of scalar-field perturbation and tensor modes.
\par 
The leading order term of the EoM for the scalar field (\ref{eq:smallscaleScalarEoM}) yields
\beqn
    \hat{g}^{ab}l_a l_b = 0,
\eeqn
where $\mathcal{B}^{(\varphi)}\neq0$ and $l_a:=-\barnabla_a \theta^\mrs$ is the wave vector of the scalar field perturbation,
\beqn
    \hat{g}^{ab}:=(\barG_{2X}+2\barX \barG_{2XX})\left(-\barn^a \barn^b + c_\text{(S)}^2 \bargamma^{ab}\right),
\eeqn
and
\beqn
    c_\text{(S)}^2:=\frac{\barG_{2X}}{\barG_{2X}+2\barX \barG_{2XX}}
\eeqn
is the sound speed squared given by the background value of the scalar field.
\par
The leading order terms of the time-space components of the metric EoMs yield also trivial conditions on the vector modes amplitudes as
\beqn
    \mathcal{A}^{\V (1,0)}_a=\mathcal{B}^{\V (1,0)}_a=0.
\eeqn
This means also that the vector modes contained in the metric are smaller than the amplitudes of the scalar-field perturbation and tensor modes.
\par
The leading order terms of the traceless part of the space-space components yield the equation for the tensor wave vector $k_a :=-\barnabla_a \theta^\T$ as
\beqn
    k^a k_a = 0,
\eeqn
where $\mathcal{A}^\T_{ab}\neq0$. This means that the tensor modes propagate at the speed of light.

\subsubsection{Next-to-leading order}
Next, we investigate the next-to-leading order. The next-to-leading order term of the time-time component of the metric EoMs yields
\beqn
    \biggl( \mathcal{A}^{\bracketC (1,1)} - i \frac{\bar{K}^{ab}}{2\omega^\T}\mathcal{A}^{\T (1,0)}_{ab} \biggr)e^{i\theta^\T/\epsilon} + \biggl( \mathcal{B}^{\bracketC (1,1)} + i \frac{ \sqrt{2\barX} \bar{G}_{2X}}{2\omega^{\mrs}{}}\mathcal{B}^{\bracketvarphi (1,0)} \biggr)e^{i\theta^\mrs/\epsilon}=0,
    \label{eq:CaseI_next-to-leading_Hamiltonian}
\eeqn
where $\omega^\T:=-k^a \barn_a $ and  $\omega^\mrs:=-l^a \barn_a $ are frequencies of the the scalar and tensor modes measured by an observer moving with four-velocity $\barn^a$. In the case where the sound speed of the scalar-field perturbation is not the speed of light i.e. $c_{\mathrm{(S)}}\neq1$, $e^{i(\theta^\mrs-\theta^\T)/\epsilon}$ oscillates rapidly over time unless the sound speed of the scalar field is fine-tuned to be close to the speed of light. Thus multiplying $e^{-i\theta^\T/\epsilon}$ by Eq. (\ref{eq:CaseI_next-to-leading_Hamiltonian}) and integrating over a time interval sufficiently longer than the oscillation period, we obtain
\beqn
    \mathcal{A}^{\bracketC (1,1)} = i \frac{\bar{K}^{ab}}{2\omega^\T}\mathcal{A}^{\T (1,0)}_{ab}.
    \label{eq:caseIevo_eq_for_next-to-leadingA_C}
\eeqn
Similarly multiplying $e^{-i\theta^\mrs/\epsilon}$ by Eq. (\ref{eq:CaseI_next-to-leading_Hamiltonian}) and integrating yield
\beqn
    \mathcal{B}^{\bracketC (1,1)}{} = - i \frac{ \sqrt{2\barX} \bar{G}_{2X}}{2\omega^{\mrs}{}}\mathcal{B}^{\bracketvarphi (1,0)}.
    \label{eq:caseIevo_eq_for_next-to-leadingB_C}
\eeqn
In the case where the scalar sound speed is the speed of light$c_{\mathrm{(S)}} = 1$, the amplitudes cannot be separated by the difference in the speed of sound, and the sum of the amplitudes satisfies the constraint
\beqn
    \mathcal{A}^{\bracketC (1,1)} + \mathcal{B}^{\bracketC (1,1)} = i \frac{\bar{K}^{ab}}{2\omega^\T}\mathcal{A}^{\T (1,0)}_{ab} - i\frac{ \sqrt{2\barX} \bar{G}_{2X}}{2\omega^{\mrs}{}}\mathcal{B}^{\bracketvarphi (1,0)}.
\eeqn
Hereafter, we write equations only for the case of $c_{\mathrm{(S)}} \neq 1$. 
Using the above results and the next-to-leading order term of the trace part of the space-space components, we obtain
\beqn
    \mathcal{A}^{\bracketA (1,1)} &=& i \frac{\bar{K}^{ab}}{2\omega^\T}\mathcal{A}^{\T (1,0)}_{ab},     \label{eq:caseIevo_eq_for_next-to-leadingA_A}\\
    \mathcal{B}^{\bracketA (1,1)}{} &=& - \frac{i \sqrt{2\barX} \bar{G}_{2X}}{2\omega^{\mrs}{}}\mathcal{B}^{\bracketvarphi (1,0)}.
        \label{eq:caseIevo_eq_for_next-to-leadingB_A}
\eeqn
These equations mean that the scalar-mode amplitudes included in the metric, which are one order higher than the tensor-mode amplitudes and the amplitude of the scalar-field perturbation, are coupled with the tensor-mode amplitudes as well as the scalar-mode amplitudes of the scalar-field perturbation.
\par
The next-to-leading order term of the EoM for the scalar field yields the evolution equation for the leading amplitude of the scalar-field perturbation $B^{\bracketvarphi (1,0)}$,
\beqn
    \barnabla_a \Bigl( \hat{g}^{ab} \bigl(\mathcal{B}^{\bracketvarphi (1,0)}\bigr)^2 l_b \Bigr)=0.
    \label{eq:caseIevo_eq_for_leadingvarphi_varphi}
\eeqn
This means that the current of the scalar field perturbation conserves in time.
\par
Using the above results and the next-to-leading order term of the time-space components of the metric EoMs, we obtain
\beqn
    \mathcal{A}^{\V (1,1)}_a &=& -i\frac{ \bar{a}^{b} \omega^{\T}{} 
    + 2 \bar{K}_{c}{}^{b} k^{c} }{\omega^{\T}{}^2}\mathcal{A}^{\T (1,0)}_{ab}, \label{eq:caseIevo_eq_for_next-to-leadingA_V}\\
    \mathcal{B}^{\V (1,1)}_{a} &=& 0.
\eeqn
This means that there are vector modes that are one order higher than the tensor modes that couple to the amplitude of the tensor modes and scalar-field perturbation.
\par
The next-to-leading order term of the traceless part of the space-space components of the metric EoMs yields the evolution of the tensor-mode amplitudes
\beqn
    0 &=&k^{c} \bar{\nabla}_{c}\mathcal{A}^{\T (1,0)}_{ab} + \frac{1}{2} \mathcal{A}^{\T (1,0)}_{ab} \bar{\nabla}_{c}k^{c} - \frac{1}{\omega^\T}\left(2\mathcal{A}^{\T (1,0)}_{d(a}k_{b)}k^c \barnabla_c \barn^d\right). 
    \label{eq:caseIevo_eq_for_Tleading_T}
\eeqn
This equation represents the conservation of the graviton number. We show that by dividing the leading amplitude $\mathcal{A}^{\T (1,0)}_{ab}$ into two independent polarizations as
\beqn
    \mathcal{A}^{\T (1,0)}_{ab}=\mathcal{A}^{\T (1,0)}_{+}e^+_{ab}+\mathcal{A}^{\T (1,0)}_{\times}e^\times_{ab},
\eeqn
where $e^{+}_{ab}e_{+}^{ab}=e^{\times}_{ab}e_{\times}^{ab}=1$ and $e^{+}_{ab}e_{\times}^{ab}=0$.
Multiplying Eq. (\ref{eq:caseIevo_eq_for_Tleading_T}) by $e^{ab}_{+}$, we obtain
\beqn
    0=k^c \barnabla_c \mathcal{A}^{\T (1,0)}_+ + \frac{1}{2}\mathcal{A}^{\T (1,0)}_+ \barnabla_c k^c .
    \label{eq:CaseIplusmode}
\eeqn
where we have used $k^a e_{ab}=0+\mathcal{O}(\epsilon)$ derived from the transverse condition $\barD^a E^\T_{ab}=0$ and $e_+^{ab} \barnabla_c e^+_{ab}=0$ derived from the normalization of the polarization tensor $e^{+}_{ab}e_{+}^{ab}=1$. Similarly, multiplying Eq. (\ref{eq:caseIevo_eq_for_Tleading_T}) by $e^{ab}_{\times}$, we obtain
\beqn
    0=k^c \barnabla_c \mathcal{A}^{\T (1,0)}_\times + \frac{1}{2}\mathcal{A}^{\T (1,0)}_\times \barnabla_c k^c .
    \label{eq:CaseIcrossmode}
\eeqn
Using Eqs. (\ref{eq:CaseIplusmode}) and (\ref{eq:CaseIcrossmode}), we obtain the following equation, which is equivalent to the conservation of the graviton number.
\beqn
    \barnabla_a \left((\mathcal{A}^{\T (1,0)})^2 k^a \right)=0,
\eeqn
where $\left(\mathcal{A}^{\T (1,0)}\right)^2:=\left(\mathcal{A}^{\T (1,0)}_+\right)^2+\left(\mathcal{A}^{\T (1,0)}_\times\right)^2$.
\par
The leading amplitudes of the tensor modes $E^\T_{ab}$ and scalar -field perturbation $\varphi$ evolve along the wave vectors, which are given in Eq. (\ref{eq:caseIevo_eq_for_Tleading_T}) and (\ref{eq:caseIevo_eq_for_leadingvarphi_varphi}), respectively. The leading amplitudes of the other perturbations $A, C, B_a^\V$ are obtained by the relation between these amplitudes and the amplitudes of $E^\T_{ab}$ and $\varphi$, Eq. (\ref{eq:caseIevo_eq_for_next-to-leadingA_A}), (\ref{eq:caseIevo_eq_for_next-to-leadingA_C}), (\ref{eq:caseIevo_eq_for_next-to-leadingA_V}), (\ref{eq:caseIevo_eq_for_next-to-leadingB_A}), and (\ref{eq:caseIevo_eq_for_next-to-leadingB_C}). Table \ref{tb:eqs_for_ampli_in_CaseI} summarizes these equations.
\par
\begin{table}[htbp]
    \centering
    \begin{tabular}{c|c||c|c||c|c}\hline
        mode & m & $\displaystyle\mathcal{A}^{(\mathrm{m})(1,0)}$ & $\mathcal{A}^{(\mathrm{m})(1,1)}$ & $\mathcal{B}^{(\mathrm{m})(1,0)}$ & $\mathcal{B}^{(\mathrm{m})(1,1)}$  \\ \hline
        & $\varphi$ & $0$ & $0$ & equation (\ref{eq:caseIevo_eq_for_leadingvarphi_varphi}) &  \\ 
        Scalar & $A$ & $0$ & equation (\ref{eq:caseIevo_eq_for_next-to-leadingA_A}) & 0 & equation (\ref{eq:caseIevo_eq_for_next-to-leadingB_A}) \\ 
        & $C$ & $0$ & equation (\ref{eq:caseIevo_eq_for_next-to-leadingA_C}) & $0$ & equation (\ref{eq:caseIevo_eq_for_next-to-leadingB_C}) \\ \hdashline
        Vector & $\text{V}$ & $0$ & equation (\ref{eq:caseIevo_eq_for_next-to-leadingA_V}) & $0$ & $0$ \\ \hdashline
        Tensor & $\text{T}$ & equation (\ref{eq:caseIevo_eq_for_Tleading_T}) & & $0$ & $0$ \\ \hline
    \end{tabular}
    \caption{Equations for the amplitudes in the case with $h^\text{(scalar)}_{ab}\sim\epsilon\varphi$}
    \label{tb:eqs_for_ampli_in_CaseI}
\end{table}

\subsection{Case with $h^\mathrm{(scalar)}_{ab}\sim\varphi$: kinetic gravity branding or/and non-minimal coupling}\label{sec:caseII}
We investigate the leading and next-to-leading order terms of Eq. (\ref{eq:smallscaleEinstein}) and (\ref{eq:smallscaleScalarEoM}) for the models with kinetic gravity branding ($G_{3X}\neq0$) or/and non-minimal coupling ($G_{4\phi}\neq 0$) in which the Lagrangian density is Eq. (\ref{eq:LagrangianCaseI}).

\subsubsection{Leading order}
We compute the leading order. As in the case with $h^\text{(scalar)}_{ab}\sim\epsilon\varphi$, the leading order terms of the time-time component of the metric EoMs and the trace part of the space-space components yield
\beqn
    \mathcal{A}^{\bracketA (1,0)}=0,\ \mathcal{A}^{\bracketC (1,0)}=0,
\eeqn
and
\beqn
    \mathcal{B}^{\bracketC (1,0)}{} &=& 
    - \frac{ \bar{G}_{4\phi} + \bar{X} \bar{G}_{3X}}{\bar{G}_{4}} \mathcal{B}^{\bracketvarphi (1,0)}{} , \label{eq:caseII_eq_for_Sleading_C}\\
    \mathcal{B}^{\bracketA (1,0)}{} &=& \frac{ \bar{G}_{4\phi} -  \bar{X} \bar{G}_{3X}}{\bar{G}_{4}} \mathcal{B}^{\bracketvarphi (1,0)}{},\label{eq:caseII_eq_for_Sleading_A}.
\eeqn
This means that there are scalar modes included in the metric that are of the same order as the scalar-field perturbation. This is a significant difference from the Case with $h^\text{(scalar)}_{ab}\sim \epsilon\varphi$.
\par
The leading order term of the EoM for the scalar field yields the equation for the wave vector as
\beqn
    \sum_{i=G_2,G_3,G_4}\Bigl[\mathcal{E}^{(i)}\Bigr]_{\mathcal{O}(h\epsilon^{-2})} 
    = \sum_{i=G_2,G_3,G_4}\hat{g}^{ab}_{(i)}l_a l_b  \mathcal{B}^{\bracketvarphi (1,0)}e^{i\theta^\mrs/\epsilon}
    =0,
    \label{eq:caseII_evo_eq_for_scalar}
\eeqn
where
\beqn
    \hat{g}^{ab}_{(G_2)}=(\barG_{2X}+2\barX \barG_{2XX})(-\barn^a \barn^b) +\barG_{2X}\bargamma^{ab},
\eeqn
\beqn
    \hat{g}^{ab}_{(G_3)}&=&\Bigl(2\barG_{3\phi}+2\barG_{3X}\bar{\Box} \barphi+2\barX \barG_{3\phi X}+\barG_{3XX}(\barnabla_c \barphi \barnabla^c \barX +2\barX\bar{\Box}\barphi)\nonumber \\
    &&+\frac{1}{\barG_{4}}(3\barX\barG_{3X}\barG_{4\phi}+3\barX^2\barG_{3X}^2)\Bigr)(-\barn^a \barn^b)\nonumber \\
    && +\Bigl(2\barG_{3\barphi}+2\barG_{3X}\bar{\Box} \barphi-2\barX \barG_{3\phi X}+\barG_{3XX}\barnabla_c \barphi \barnabla^c X \nonumber \\
    && +\frac{1}{\barG_{4}}(\barX \barG_{3X}\barG_{4\phi}-\barX^2 \barG^2_{3X})\Bigr)\bargamma^{ab}\nonumber \\
    &&+(-2\barG_{3XX}\barnabla^{(a} \barX \barnabla^{b)} \barphi - 2\barG_{3X}\barnabla^{(a} \barnabla^{b)} \barphi),
\eeqn
\beqn
    \hat{g}^{ab}_{(G_4)}=\frac{1}{\barG_4}\Bigl( (3\barG_{4\phi}^2+3\barX \barG_{3X}\barG_{4\phi})(-\barn^a \barn^b)  +(3\barG_{4\phi}^2+\barX\barG_{3X}\barG_{4\phi})\bargamma^{ab}\Bigr).
\eeqn
For convenience of notation, we write $\hat{g}^{ab} = \hat{g}^{ab}_{(G_2)}+\hat{g}^{ab}_{(G_3)}+\hat{g}^{ab}_{(G_4)}$. Then Eq. (\ref{eq:caseII_evo_eq_for_scalar}) yields
\beqn
    \hat{g}^{ab}l_a l_b=0,
    \label{eq:CaseII_eq_for_soundspeed}
\eeqn
where $\mathcal{B}^{(\varphi)}\neq0$.
For $G_{3X}\neq 0$, the scalar field is regarded as an imperfect fluid \cite{Deffayet:2010qz}, thus the sound speed depends on the propagation direction and the background value of the scalar field.
\par
The leading order terms of the time-space components of the metric EoMs yield the condition on the vector-mode amplitudes  as
\beqn
    \mathcal{A}^{\V (1,0)}_a=0,\ \mathcal{B}^{\V (1,0)}_{a}&=&0\label{eq:caseII_eq_for_Sleading_V}.
\eeqn
This means that the vector modes contained in the metric are smaller than the amplitudes of the scalar field perturbation and tensor modes as in the previous case.
\par
The traceless part for the space-space components of the metric EoMs yields that the wave vector is null, i.e.
\beqn
    k^a k_a = 0,
\eeqn
where $\mathcal{A}^\T_{ab}\neq0$.

\subsubsection{Next-to-leading order}
We now compute the next-to-leading order. The next-to-leading order terms of the time-time component of the metric EoMs and the trace part of the space-space components yield the conditions for the scalar-mode amplitudes as
\beqn
    \mathcal{A}^{\bracketvarphi (1,1)}{} &=& 0,\\
    \mathcal{A}^{\bracketC (1,1)}{} &=& \mathcal{A}^{\bracketA (1,1)}{} = i\frac{\barK^{ab}}{2\omega^\T}\mathcal{A}^{\T (1,0)}_{ab}. \label{eq:caseII_eq_for_Tleading_AC}
\eeqn
These equations mean that scalar and tensor modes do not decouple from each other at this order.
Note that the derivative of the factors of $\mathcal{B}^{\bracketvarphi (1,0)}{}$ in the equation (\ref{eq:caseII_eq_for_Sleading_A}) and (\ref{eq:caseII_eq_for_Sleading_C}) should be taken into account when calculating the conditions for the next-to-leading order amplitudes $\mathcal{B}^{(A)(1,1)},\mathcal{B}^{(C)(1,1)}$.
Those conditions can be written using the amplitudes one order lower, and each mode does not decouple at this order.
Those equations are not written in this paper because it is lengthy for presentation.
\par
Using Eq. (\ref{eq:CaseII_eq_for_soundspeed})-(\ref{eq:caseII_eq_for_Sleading_V}) and the next-to-leading order term of EoM for the scalar field, we obtain the evolution equation for the leading amplitude of the scalar-field perturbation
\beqn
    &&\barnabla_a\left(\hat{g}^{ab} l_b \left(\mathcal{B}^{\bracketvarphi (1,0)}\right)^2\right)=\nn\\
    &&\frac{-\left(\mathcal{B}^{\bracketvarphi (1,0)}\right)^2\bar{G}_{3X} }{\barG_{4}^2}\Bigl[ 
    \Bigl(4 \bar{G}_{4} \bar{X}^2 \bar{a}^a l_{a} 
    - \omega^{\mrs} \bigl(
    -6 \bar{K} \bar{G}_{4} \bar{X}^2 
    - 3 \bar{G}_{4} \bar{X} \bar{n}^{b} \bar{\nabla}_{b}\bar{X} 
    - 6 \sqrt{2} \bar{X}^{5/2} \bar{G}_{4\phi}\bigr)\Bigr) \bigl(\bar{G}_{3X}\bigr) \nn \\
    &&+ 4 \bar{G}_{4} \bar{X}^3 \bar{a}^a l_{a} \bar{G}_{3XX} 
    - \omega^{\mrs} \Bigl(
    -6 \sqrt{2} \bar{X}^{3/2} \bigl(\bar{G}_{4\phi}\bigr)^2 
    - 2 \sqrt{2} \bar{G}_{4} \bar{X}^{3/2} \bar{G}_{2X} 
    -  \sqrt{2} \bar{G}_{4} \bar{X}^{5/2} \bar{G}_{2XX} 
    - 2 \bar{K} \bar{G}_{4} \bar{X}^3 \bar{G}_{3XX} \nn \\
    &&- 3 \bar{G}_{4} \bar{X}^2 \bar{n}^{b} \bar{\nabla}_{b}\bar{X} \bar{G}_{3XX} 
    - 4 \sqrt{2} \bar{G}_{4} \bar{X}^{3/2} \bar{G}_{3\phi} 
    + 2 \sqrt{2} \bar{G}_{4} \bar{X}^{5/2} \bar{G}_{3\phi X}\Bigr)
    \Bigr].
    \label{eq:caseII_evo_eq_for_leading_varphi}
\eeqn
The right-hand side of this equation is zero when kinetic gravity branding is absent, i.e. $G_{3X}=0$. Then the current for the scalar field perturbation conserves in time.
The fact that the current does not conserve for the model with kinetic gravity branding is consistent with the fact that the fluid has the diffusion when this model is interpreted as an imperfect fluid \cite{pujolas2011imperfect}.
\par
The next-to-leading order terms of the time-space components of the metric EoMs yield the conditions for the vector-mode amplitudes as
\beqn
    \mathcal{A}^{\V (1,1)}_{a} &=& -i\frac{ \bar{a}^{b} \omega^{\T}{} 
    + 2 \bar{K}_{c}{}^{b} k^{c} }{\omega^{\T}{}^2}\mathcal{A}^{\T (1,0)}_{ab} .\label{eq:caseII_eq_for_Tleading_V}
\eeqn
This means that the vector-mode amplitudes $\mathcal{A}^{\V (1,1)}_{a}$ couple with the tensor-mode amplitudes at one order lower.
The equation for $ \mathcal{B}^{\V (1,1)}_{a}$ shows that the vector-mode amplitudes $\mathcal{B}^{\V (1,1)}_{a}$ couple with the scalar-mode amplitudes at one order lower. It is not written in this paper because it is lengthy for presentation.
\par
The next-to-leading order term of the traceless part of the space-space components of the metric EoMs yields the evolution equation for the tensor-mode leading amplitude,
\beqn
    k^{c} \bar{\nabla}_{c}\mathcal{A}^{\T (1,0)}_{ab} + \frac{1}{2} \mathcal{A}^{\T (1,0)}_{ab} \bar{\nabla}_{c}k^{c} - \frac{1}{\omega^\T}\left(2\mathcal{A}^{\T (1,0)}_{d(a}k_{b)}k^c \barnabla_c \barn^d\right)- \frac{ \sqrt{2\bar{X}} \omega^{\T}{}  \bar{G}_{4\phi}}{2 \bar{G}_{4}}\mathcal{A}^{\T (1,0)}_{ab} = 0
    \label{eq:CaseIItensorleadingamplitude}
\eeqn
and the equation for $\mathcal{B}^{\T (1,1)}_{ab}$,
\beqn
    l_{a} l^{a} \mathcal{B}^{\T (1,1)}_{cd} 
    - \frac{4i \omega^{\mrs} \mathcal{B}^{\bracketvarphi (1,0)} \barX \bar{G}_{3X}}{3\barG_4}\Bigl(3\bar{K}_{cd}  
    - \bar{K} \bar{g}_{cd}
    -  \bar{K} \bar{n}_{c} \bar{n}_{d} \Bigr) = 0.
    \label{eq:caseII_traceless_next-to-leading_T}
\eeqn
This also means that each mode does not decouple at this order. By the same procedure as in the previous case, Eq. (\ref{eq:CaseIItensorleadingamplitude}) yields the following equation, which implies conservation of the graviton number,
\beqn
    \frac{1}{2\barG_{4}}\barnabla_c \left(\barG_{4}(\mathcal{A}^{\T (1,0)})^2 k^c \right)=0.
    \label{eq:caseIIevo_eq_for_Tleading_T}
\eeqn
The graviton flux density $4$-vector in the subclass we consider is $(\barG_{4}(\mathcal{A}^{\T (1,0)}_{ab})^2 k^c )/(4\pi G_\mathrm{N} \hbar)$ \cite{Dalang:2019rke}. Thus this equation expresses graviton number conservation.
\par
As in the case with $h^\text{(scalar)}_{ab}\sim\epsilon\varphi$, the evolution equations for the leading amplitudes along the wave vectors of the tensor mode $E^\T_{ab}$ and scalar field perturbation $\varphi$ are Eq. (\ref{eq:caseIIevo_eq_for_Tleading_T}) and (\ref{eq:caseII_evo_eq_for_leading_varphi}), respectively.
The leading amplitudes of the other perturbations $A, C, B_a^\V$ are obtained by the relation between these amplitudes and the amplitudes of $E^\T$ and $\varphi$, Eq. (\ref{eq:caseII_eq_for_Tleading_AC}), (\ref{eq:caseII_eq_for_Tleading_V}), (\ref{eq:caseII_eq_for_Sleading_A}) , and (\ref{eq:caseII_eq_for_Sleading_C}).
Table \ref{tb:eqs_for_ampli_in_CaseII} summarizes these equations.
\begin{table}[htbp]
    \centering
    \begin{tabular}{c|c||c|c||c|c}\hline
        mode & m & $\displaystyle\mathcal{A}^{(\mathrm{m})(1,0)}$ & $\mathcal{A}^{(\mathrm{m})(1,1)}$ & $\mathcal{B}^{(\mathrm{m})(1,0)}$ & $\mathcal{B}^{(\mathrm{m})(1,1)}$  \\ \hline
        & $\varphi$ & $0$ & $0$ & equation (\ref{eq:caseII_evo_eq_for_leading_varphi}) &  \\ 
        Scalar & $A$ & $0$ & equation (\ref{eq:caseII_eq_for_Tleading_AC}) & equation (\ref{eq:caseII_eq_for_Sleading_A}) & \\ 
        & $C$ & $0$ & equation (\ref{eq:caseII_eq_for_Tleading_AC}) & equation (\ref{eq:caseII_eq_for_Sleading_C}) &  \\ \hdashline
        Vector &$\text{V}$ & $0$ & equation (\ref{eq:caseII_eq_for_Tleading_V}) & $0$ &  \\ \hdashline
        Tensor & $\text{T}$ & equation (\ref{eq:caseIIevo_eq_for_Tleading_T}) & & $0$ & equation (\ref{eq:caseII_traceless_next-to-leading_T}) \\ \hline
    \end{tabular}
    \caption{Equations for the amplitudes in the case with $h^\text{(scalar)}_{ab}\sim\varphi$}
    \label{tb:eqs_for_ampli_in_CaseII}
\end{table}

\subsection{Effective metric in Generalized Brans-Dicke theories}
In Generalized Brans-Dicke theories ($G_{3}=0$), the high-frequency part of the EoM for the scalar field and the traceless part of the space-space components of the metric EoMs can be concisely written as
\beqn
    \biggl[\Bigl(\hat{g}^{ab}l_a l_b (\mathcal{B}^{\bracketvarphi(1,0)})^2\Bigr) h\epsilon^{-2} + \Bigl( \barnabla_a \bigl(\hat{g}^{ab}(\mathcal{B}^{\bracketvarphi(1,0)})^2 l_b\bigr)\Bigr)i h\epsilon^{-1} + \mathcal{O}(h,h^2\epsilon^{-2}) \biggr]e^{i\theta^\mrs/\epsilon} + \biggl[\mathcal{O}(h,h^2\epsilon^{-2}) \biggr]e^{i\theta^\T/\epsilon}=0,
\eeqn
and
\beqn
    \biggl[\Bigl(\barG_{4}\barg^{ab} k_a k_b (\mathcal{A}^{\T (1,0)})^2\Bigr)  h\epsilon^{-2} + \Bigl( \bar{\nabla}_a\bigl(\barG_{4}\barg^{ab} (\mathcal{A}^{\T (1,0)})^2 k_b\bigr)\Bigr)i h\epsilon^{-1} + \mathcal{O}(h,h^2\epsilon^{-2})\biggr]e^{i\theta^\T/\epsilon} \nn \\
    + \biggl[\mathcal{O}(h,h^2\epsilon^{-2})\biggr]e^{i\theta^\mrs/\epsilon}=0.
\eeqn
For both modes, the high-frequency part of the EoM can be written in the same form
\beqn
    \Biggl[\biggl(\hat{\mathscr{G}}^{ab}\mathscr{K}_a \mathscr{K}_b \mathscr{A}^2\biggr) h\epsilon^{-2} + \biggl( \barnabla_a \Bigl(\hat{\mathscr{G}}^{ab}\mathscr{K}_b \mathscr{A}^2 \Bigr)\biggr)i h\epsilon^{-1} + \mathcal{O}(h,h^2\epsilon^{-2}) \Biggr]e^{i\theta/\epsilon} + \Biggl[ \mathcal{O}(h,h^2\epsilon^{-2}) \Biggr]e^{i\theta'/\epsilon}=0.
    \label{eq:evolution}
\eeqn
Here, $\hat{\mathscr{G}}_{ab}$ is a precursor metric in terms of which the perturbation move along null rays as $\hat{\mathscr{G}}^{ab}\mathscr{K}_a \mathscr{K}_b=0$ and $\barnabla_a \left(\hat{\mathscr{G}}^{ab}\mathscr{K}_b \mathscr{A}^2 \right)=0$, where $\mathscr{A}$ is the leading amplitude, $\mathscr{K}_a:=-\barnabla_a\theta$ is the wave vector, and $\theta'$ denote the other phase.

The equation (\ref{eq:evolution}) becomes clearer if it is rewritten in terms of an effective metric $\mathscr{G}^\text{(eff)}_{ab}$ \footnote{While in the present setup tensor modes are luminal, in more general setups the effective metric can be defined for nonluminal tensor modes. For example, \cite{Hajian:2020dcq} introduces the effective metric on which tensor modes propagate along null rays in order to define a black hole temperature that is appropriate for black hole thermodynamics.} and the associated covariant derivative operator $\nabla^\text{(eff)}_a$, which satisfies the metric compatibility $\nabla^\text{(eff)}_c \mathscr{G}^\text{(eff)}_{ab}=0$.
We thus look for an effective metric $\mathscr{G}^\text{(eff)}_{ab}$ that satisfies the following conditions. 
\begin{enumerate}[(a)]
    \item The perturbation should move on the null ray of the effective metric, i.e. $\mathscr{G}_\text{(eff)}^{ab}\mathscr{K}_a\mathscr{K}_b = 0$.
    \item The current conservation \footnote{For photons propagating in a spacetime with the metric $g_{ab}$, the current conservation of the number of photons can be written as $g^{ab} \nabla_a \left(k_b \mathcal{A}^2\right)=0$, where $k^a$ is the wave vector and $\mathcal{A}$ is the amplitude.} can be rewritten in the form $\mathscr{G}_\text{(eff)}^{ab} \nabla^\text{(eff)}_a \left(\mathscr{K}_b \mathscr{A}^2\right)=0$.
\end{enumerate}
In this section, we derive the effective metric satisfying the above two conditions.
\par
Since the perturbations propagate along null rays of the precursor metric $\hat{\mathscr{G}}_{ab}$, the effective metric should be obtained by a conformal transformation of $\hat{\mathscr{G}}_{ab}$, that does not change the null directions, as
\beqn
    {\mathscr{G}}^{ab}_\text{(eff)}
    =\mathscr{X}\hat{\mathscr{G}}^{ab}
    =\mathscr{X}\left(-\mathscr{Y}\barn^a \barn^b + \mathscr{Z} \bargamma^{ab}\right).\label{eq:effectivemetric}
\eeqn
where $\mathscr{X}$, $\mathscr{Y}$, and $\mathscr{Z}$ are functions of $\barphi$ and $\barX$.
Thus the effective metric is written as
\beqn
    \mathscr{G}^\text{(eff)}_{ab}=
    \frac{1}{\mathscr{X}} \left(-\frac{1}{\mathscr{Y}}\barn^a\barn^b + \frac{1}{\mathscr{Z}}\bargamma^{ab}  \right),
\eeqn
where $\mathscr{G}^\text{(eff)}_{ab}$ is defined to satisfy $\mathscr{G}_\text{(eff)}^{ac}\mathscr{G}^\text{(eff)}_{cb}=\delta^a_{\ b}$\footnote{The indices of the effective metric cannot be raised and lowered by $\barg_{ab}$, i.e., $\mathscr{G}^\text{(eff)}_{ab} \neq \barg_{ac}\barg_{bd} \mathscr{G}_\text{(eff)}^{cd} $. The indices can be raised and lowered by the effective metric itself and its inverse.}. 
The determinant of the effective metric can be written as 
\beqn
    \det\left(\mathscr{G}^\text{(eff)}_{ab}\right)=(\mathscr{X}^4\mathscr{Y}\mathscr{Z}^3)^{-1}\det( \barg_{ab}).
\eeqn

By the condition (b), the effective metric should satisfy
\beqn
   \mathscr{G}_\text{(eff)}^{ab}\nabla^\text{(eff)}_a(\mathscr{K}_b \mathscr{A}^2)
   =\frac{1}{\sqrt{-\det\left(\mathscr{G}^\text{(eff)}_{cd}\right)}}\partial_a \left(\sqrt{-\det\left(\mathscr{G}^\text{(eff)}_{cd}\right)}\mathscr{G}_\text{(eff)}^{ab} \mathscr{K}_b \mathscr{A}^2 \right)=0
\eeqn
 (See e.g. Eq. (3.4.10) of \cite{wald2010general}). The evolution equation for the amplitude is 
\beqn
    \barnabla_a \left(\hat{\mathscr{G}}^{ab} \mathscr{K}_b \mathscr{A}^2 \right)
    =\frac{1}{\sqrt{-\det(\barg_{cd})}}\partial_a \left(\sqrt{-\det(\barg_{cd})}\hat{\mathscr{G}}^{ab} \mathscr{K}_b \mathscr{A}^2 \right)=0,
\eeqn
By comparing the above two equations, we can see that the function $\mathscr{X}$ should be chosen to satisfy 
\beqn
    \sqrt{-\det\left(\mathscr{G}^\text{(eff)}_{cd}\right)}\mathscr{G}^{ab}_\text{(eff)}=\sqrt{-\det(\barg_{cd})}\hat{\mathscr{G}}^{ab}.
    \label{eq:determinatcondition}
\eeqn
This equation yields
\beqn
    \mathscr{X}=\frac{1}{\sqrt{\mathscr{Y}\mathscr{Z}^3}}=\frac{\mathscr{C}}{\mathscr{Z}^2}
\eeqn
where $\mathscr{C}:=\sqrt{\mathscr{Z}/\mathscr{Y}}$ is the sound speed. We conclude that the effective metric is given by
\beqn
    \mathscr{G}_\text{(eff)}^{ab}=\frac{1}{\mathscr{Z}\mathscr{C}}\left(-\barn^a\barn^b + \mathscr{C}^{2}\bargamma^{ab}\right),\  \mathscr{G}^\text{(eff)}_{ab}=\mathscr{Z}\mathscr{C}\left(-\barn_a\barn_b + \mathscr{C}^{-2}\bargamma_{ab}\right).
\eeqn
\par
Using this effective metric, Eq. (\ref{eq:evolution}) can be rewritten in the form
\beqn
    \Biggl[\biggl({\mathscr{G}}_\text{(eff)}^{ab}\mathscr{K}_a \mathscr{K}_b \mathscr{A}^2\biggr) h\epsilon^{-2} + \biggl(\mathscr{G}_\text{(eff)}^{ab} \nabla^\text{(eff)}_a \Bigl(\mathscr{K}_b \mathscr{A}^2 \Bigr)\biggr)i h\epsilon^{-1} + \mathcal{O}(h,h^2\epsilon^{-2}) \Biggr]e^{i\theta/\epsilon} + \Biggl[\mathcal{O}(h,h^2\epsilon^{-2}) \Biggr]e^{i\theta'/\epsilon}=0.
\eeqn
Furthermore, this equation can be written more simply by using the effective d'Alembert operator $ \Box^\text{(eff)}:={\mathscr{G}}_\text{(eff)}^{ab}\nabla^\text{(eff)}_a \nabla^\text{(eff)}_b$ as
\beqn
    \Box^\text{(eff)}\left(\mathscr{A}e^{i\theta/\epsilon}\right)+\mathcal{O}(h,h^2\epsilon^{-2})=0.
    \label{eq:effective_dalembert}
\eeqn
The effective metrics $\barg_{(\mathrm{T},\mathrm{S})}^{ab}$ for the tensor modes and scalar field perturbation are written as
\begin{numcases}{}
    \barg_{(\mathrm{T})}^{ab}= \barg_{(\mathrm{E})}^{ab}=\frac{1}{\barG_4} \barg^{ab} & for $E_{ab}^\T$,\\
    \barg_{(\mathrm{S})}^{ab}= \frac{\barG_{4}\sqrt{\barG_{2X}{\barG_{4}}+2\barX \barG_{2XX}{\barG_{4}}+3\barG_{4\phi}^2}}{\sqrt{\left(\barG_{2X}{\barG_{4}}+3\barG_{4\phi}^2\right)^3}}\left(-\barn^a \barn^b +\frac{\barG_{2X}{\barG_{4}}+{3\barG_{4\phi}^2}}{\barG_{2X}{\barG_{4}}+2\barX \barG_{2XX}{\barG_{4}}+3\barG_{4\phi}^2} \bargamma^{ab}\right) & for $\varphi$.
\end{numcases}
where ${\barg}_{\mathrm{(E)},ab}$ is the metric in the Einstein frame. 
The effective metric for tensor modes is a metric in the Einstein frame, in which the conservation equation of the graviton in GR obviously holds. 

\subsection{Comparison with previous studies}
In \cite{garoffolo2020gravitational}, the authors derived the evolution equations of the amplitudes for a subclass of theories in which the sound speeds of not only tensor modes but also the scalar field are equal to the speed of light. 
Later, in \cite{dalang2021scalar}, the authors derived the equations not only for the subclass in which the sound speed of the scalar field is equal to the speed of light, but also for a subclass in which it is not equal.
In this subsection, we compare our study with the latter.
\par
In \S \ref{sec:caseI}, we have studied the case which simply adds the k-essence term, i.e. the case with $h^\text{(scalar)}_{ab}\sim\epsilon\varphi$, which was not studied in \cite{dalang2021scalar}. The evolution of the amplitudes of the tensor modes is the same as in GR. The evolution of the amplitude of the scalar field perturbation is obtained by Eq. (\ref{eq:caseIevo_eq_for_leadingvarphi_varphi}), and the amplitudes of the scalar and vector modes included in metric perturbations are obtained by the relation between these and the amplitudes of the scalar-field perturbation and the tensor modes, (\ref{eq:caseIevo_eq_for_next-to-leadingA_C}), (\ref{eq:caseIevo_eq_for_next-to-leadingB_C}), (\ref{eq:caseIevo_eq_for_next-to-leadingA_A}), and (\ref{eq:caseIevo_eq_for_next-to-leadingB_A}).
\par
In \S \ref{sec:caseII}, we have studied the case with the kinetic gravity branding $G_{3}(\phi,X)\Box\phi$ or/and the nonminimal coupling $G_{4}(\phi)R$, i.e. the case with $h^\text{(scalar)}_{ab}\sim\varphi$, which was well studied in \cite{dalang2021scalar}. The evolution of the tensor modes amplitudes is given by Eq. (\ref{eq:caseIIevo_eq_for_Tleading_T}). Only when there is nonminimal coupling $G_{4}(\phi)R$, the evolution differs from GR. This is known as ``Running Planck mass'' \cite{Lagos:2019kds} on a FLRW background. The evolution of the scalar field perturbation amplitude is given by Eq. (\ref{eq:caseII_evo_eq_for_leading_varphi}) and the amplitudes of the scalar modes included in metric perturbations are obtained by the relation between them and the amplitude of the scalar-field perturbation, (\ref{eq:caseII_eq_for_Sleading_A}) and (\ref{eq:caseII_eq_for_Sleading_C}). These results are consistent with \cite{dalang2021scalar}. The authors of \cite{dalang2021scalar} extracted the tensor modes from the metric perturbation $h_{ab}$ by defining 
\beqn
    h^{(\text{tensor})}_{ab}:=\biggl(h_{ab}-\frac{1}{2}\barg_{ab}\barg^{cd}h_{cd}\biggr)+\frac{\barG_{3X}\barnabla_a \barphi \barnabla_b \barphi -\barG_{4\phi}\barg_{ab}}{\barG_4}\varphi,
    \label{eq:ExtractTensorMode}
\eeqn
and imposing the transverse condition $\barnabla^a h^{(\text{tensor})}_{ab}=0$. 
Then $h^{(\text{tensor})}_{ab}$ propagates at the speed of light and $\varphi$ propagates at the sound speed of the scalar field.
Distinct from such previous analysis, we used the SVT decomposition, clarifying the couplings among the scalar, vector and tensor modes. Furthermore, we have rewritten these equations for perturbative variables into the simple form Eq. (\ref{eq:effective_dalembert}) in Generalized Brans-Dicke theories by using the effective metric, which was not derived in \cite{dalang2021scalar}.

\section{Summary and Discussion}\label{ch:summary}
In this paper, we have formulated a systematic way to compute higher orders of the geometric optical expansion in the subclass (\ref{eq:LHorndeski}) of Horndeski theory, which includes major dark energy models, to study the phenomenology of the propagation of the metric and scalar perturbations over the curved background spacetime with distributions of matter and dark energy. 
\par
To develop the formulation for the propagation of GWs, we have assumed that the amplitudes and wavelengths are sufficiently small and the hierarchy (\ref{eq:hierarchy}) between the smallness parameter $h$ of the amplitudes and the shortness parameter $\epsilon$ of the wavelength of the perturbed variables holds. We needed to decompose the EoMs into equations for each mode. For this purpose, we have assumed that at the level of the background the derivative of the scalar field is non-vanishing and timelike. We then chose the contours of the background scalar field as the time slices at the level of the background. This choice of time slices is advantageous and makes it easier to understand the behavior of GWs intuitively, since the sound cones of both tensor and scalar GWs are upright with respect to the constant time hypersurfaces, i.e. the axes of sound cones are parallel with each other and normal to the constant time hypersurfaces, whenever the scalar field behaves as a perfect fluid. We have presented the way to systematically compute the higher orders in the geometrical optics and perturbative expansions and derive the leading and next-to-leading order of the high-frequency part of the EoMs. Then we were able to obtain the equations for the wave vectors of scalar and tensor perturbations and the evolution equation for their leading amplitudes in the direction of the wave vectors.
\par
We have studied only the leading and next-to-leading order of EoMs. When the wavelengths of perturbations approach the curvature radius of the background spacetime, geometrical optics approximation is no longer available, and the higher-order corrections are not negligible. To study the next-to-next-to-leading order, which has not been solved in the present paper, we need further details of the hierarchy between $h$ and $\epsilon$ beyond (\ref{eq:hierarchy}), i.e. which is larger, $h$ or $\epsilon^2$. Assuming the hierarchy, we can obtain the equation for the higher-order amplitudes. The highlight is that the equations for perturbed variables obey a simple wave equation equation as shown Eq. (\ref{eq:effective_dalembert}) in Generalized Brans-Dicke theories. This feature has not been found in the literature e.g. \cite{dalang2021scalar}. 
\par
Although we have focused only on GW propagation, GW generation can also contain important features of gravity and dark energy. Such features are necessary to be understood toward upcoming observational tests. For example, the orbital evolution of binary systems in Horndeski theory has been calculated in \cite{Chowdhuri:2022jyk}. In the subclasses we consider, the screening mechanism such as Vainshtein mechanism~\cite{Vainshtein:1972sx}, chameleon mechanism~\cite{Khoury:2003rn}, and k-mouflage~\cite{Babichev:2009ee} can work. In such cases, the signals of any deviation from GR may be suppressed  \cite{Renevey:2021tcz}, while it is not clearly understood how GWs change in the presence of screening. This topic remains opened for future work.

\section*{Acknowledgments}
Some calculations, such as derivation of perturbed equations of motion, were performed using a {\it Mathematica} package, xAct\footnote{\url{http://www.xact.es/}}. The work of S.M.~was supported in part by JSPS Grants-in-Aid for Scientific Research No.~17H02890, No.~17H06359, and by World Premier International Research Center Initiative (WPI), MEXT, Japan. 

\appendix
\section{The sound speeds on FLRW background in Horndeski theory}\label{app:Horndeski}
The Lagrangian density of Horndeski theory is given below using four arbitrary functions of a scalar field $\phi$ and the canonical kinetic term $X\equiv -g^{ab}\partial_a\phi\partial_b\phi/2$, 
\beqn
    \mathcal{L}= G_{2}(\phi, X)+G_{3}(\phi, X) \square \phi+G_{4}(\phi, X) R+G_{4 X}\left[(\square \phi)^{2}-\phi^{;ab} \phi_{;ab}\right] \nn \\
    +G_{5}(\phi, X) G^{ab} \phi_{;ab}-\frac{G_{5 X}}{6}\left[(\square \phi)^{3}-3 \square \phi \phi^{;ab} \phi_{;ab}+2 \phi_{;ab} \phi^{;bc} \phi_{;c}^{;a}\right],
\eeqn
where $\phi_{;ab}:=\nabla_b \nabla_a \phi$.
In \cite{Kobayashi:2011nu}, the authors derived the sound speeds of tensor modes and scalar modes on a flat FLRW background spacetime
\beqn
    \mathrm{d}s^2 = - \mathrm{d}t^2 + a^2(t) \delta_{ij}\mathrm{d}x^i \mathrm{d}x^j,
\eeqn
where $a$ is the scale factor. The squared sound speed of the tensor modes is given by 
\beqn
    c_\mathrm{(T)}^2=\frac{\mathcal{F}_\mathrm{(T)}}{\mathcal{G}_\mathrm{(T)}},
    \label{eq:soundspeedtensorperturbationonFLRW}
\eeqn
where 
\beqn
    \mathcal{F}_\mathrm{(T)} &:=&2\left[G_{4}-X\left(\ddot{\phi} G_{5 X}+G_{5 \phi}\right)\right], \\
    \mathcal{G}_\mathrm{(T)} &:=&2\left[G_{4}-2 X G_{4 X}-X\left(H \dot{\phi} G_{5 X}-G_{5 \phi}\right)\right],
\eeqn
$\dot{\phi}:=\partial_t \phi$, $\ddot{\phi} = \partial^2_t \phi$, and $H$ is the Hubble expansion rate, $H:=\partial_t a/a$. 
The squared sound speed of scalar modes is given by 
\beqn
    c_\mathrm{(S)}^2=\frac{\mathcal{F}_\mathrm{(S)}}{\mathcal{G}_\mathrm{(S)}},
    \label{eq:soundspeedscalarperturbationonFLRW}
\eeqn
where 
\beqn
    \mathcal{F}_\mathrm{(S)} &:=&\frac{1}{a} \frac{d}{d t}\left(\frac{a}{\Theta} \mathcal{G}_\mathrm{(T)}^{2}\right)-\mathcal{F}_\mathrm{(T)}, \\
    \mathcal{G}_\mathrm{(S)} &:=&\frac{\Sigma}{\Theta^{2}} \mathcal{G}_\mathrm{(T)}^{2}+3 \mathcal{G}_\mathrm{(T)},\\
    \Sigma&:=& X G_{2X}+2 X^{2} G_{2X X} - 12 H \dot{\phi} X G_{3 X} \nn \\
    &&- 6 H \dot{\phi} X^{2} G_{3 X X} + 2 X G_{3 \phi} + 2 X^{2} G_{3 \phi X}-6 H^{2} G_{4} \nn \\
    &&+6\left[H^{2}\left(7 X G_{4 X}+16 X^{2} G_{4 X X}+4 X^{3} G_{4 X X X}\right)\right. \nn \\
    &&\left.-H \dot{\phi}\left(G_{4 \phi}+5 X G_{4 \phi X}+2 X^{2} G_{4 \phi X X}\right)\right] \nn \\
    &&+30 H^{3} \dot{\phi} X G_{5 X}+26 H^{3} \dot{\phi} X^{2} G_{5 X X} \nn \\
    &&+4 H^{3} \dot{\phi} X^{3} G_{5 X X X}-6 H^{2} X\left(6 G_{5 \phi}+9 X G_{5 \phi X}+2 X^{2} G_{5 \phi X X}\right), \\
    \Theta&:=&\dot{\phi} X G_{3 X}+2 H G_{4}-8 H X G_{4 X}-8 H X^{2} G_{4 X X}+\dot{\phi} G_{4 \phi}+2 X \dot{\phi} G_{4 \phi X} \nn \\
    &&-H^{2} \dot{\phi}\left(5 X G_{5 X}+2 X^{2} G_{5 X X}\right)+2 H X\left(3 G_{5 \phi}+2 X G_{5 \phi X}\right).
\eeqn

\section{Gauge transformations}\label{gaugetrnsformations}
Under the infinitesimal gauge transformation $x^a\to x'^a=x^a+\xi^a$, the metric perturbation is transformed  as $h_{ab}\to h'_{ab}=h_{ab}-\Lie_\xi g_{ab}$, and each component of the metric perturbations is transformed as
\beqn
    A\to A'&:=&h'_{ab}\barn^a \barn^b \nn\\
    &=&A + 2 \bar{a}^{a} L^{\V}{}_{a} + 2 \bar{a}^{a} \bar{\nabla}_{a}L + 2 \bar{n}^{a} \bar{\nabla}_{a}T,\\
    B^a \to B'^a&:=& - h'_{ab}\barn^a \bargamma^{bc}\nn \\
    &=&B{}^{c} -  \bar{a}^{a} \bar{n}^{c} L^{\V}{}_{a} -  \bar{K}^{c}{}_{a} L^{\V}{}^{a} + \bar{a}^{c} T + \bar{a}^{c} \bar{n}^{a} \bar{\nabla}_{a}L \nn \\
    &&+ \bar{n}^{a} \bar{\nabla}_{a}L^{\V}{}^{c} -  \bar{K}^{c}{}_{a} \bar{\nabla}^{a}L -  \bar{\gamma}^{c}{}_{a} \bar{\nabla}^{a}T + \bar{\gamma}^{cb} \bar{n}^{a} \bar{\nabla}_{b}\bar{\nabla}_{a}L,\\
    C\to C' &:=& \frac{1}{3}h'_{ab}\bargamma^{ab} \nn\\
    &=&\frac{1}{3}\bigl(3C - 2 \bar{K} T - 2 \bar{K} \bar{n}^{a} \bar{\nabla}_{a}L - 2 \bar{\gamma}^{ab} \bar{\nabla}_{b}\bar{\nabla}_{a}L\bigr),\\
    F_{ab}\to F'_{ab} &:=& h'_{ab}\left(\bargamma^{a}_{\ c}\bargamma^{b}_{\ d}-\frac{1}{3}\bargamma^{cd}\bargamma_{ab}\right) \nn\\
    &=&F{}_{cd} + \bar{K}_{da} \bar{n}_{c} L^{\V}{}^{a} + \bar{K}_{ca} \bar{n}_{d} L^{\V}{}^{a} - 2 \bar{K}_{cd} T + \frac{2}{3} \bar{K} \bar{\gamma}_{cd} T - 2 \bar{K}_{cd} \bar{n}^{a} \bar{\nabla}_{a}L \nn\\
    && + \frac{2}{3} \bar{K} \bar{\gamma}_{cd} \bar{n}^{a} \bar{\nabla}_{a}L -  \bar{\gamma}_{db} \bar{n}^{a} \bar{n}_{c} \bar{\nabla}_{a}L^{\V}{}^{b} -  \bar{\gamma}_{cb} \bar{n}^{a} \bar{n}_{d} \bar{\nabla}_{a}L^{\V}{}^{b} + 2 \bar{n}^{a} \bar{n}^{b} \bar{n}_{c} \bar{n}_{d} \bar{\nabla}_{b}L^{\V}{}_{a} \nn \\
    &&+ \frac{2}{3} \bar{\gamma}^{ab} \bar{\gamma}_{cd} \bar{\nabla}_{b}\bar{\nabla}_{a}L - 2 \bar{n}^{a} \bar{n}^{b} \bar{n}_{c} \bar{n}_{d} \bar{\nabla}_{b}\bar{\nabla}_{a}L+ \bar{n}^{a} \bar{n}_{d} \bar{\nabla}_{c}L^{\V}{}_{a} -  \bar{\gamma}_{da} \bar{\nabla}_{c}L^{\V}{}^{a} \nn \\
    &&- 2 \bar{n}^{a} \bar{n}_{d} \bar{\nabla}_{c}\bar{\nabla}_{a}L + \bar{n}^{a} \bar{n}_{c} \bar{\nabla}_{d}L^{\V}{}_{a} -  \bar{\gamma}_{ca} \bar{\nabla}_{d}L^{\V}{}^{a} - 2 \bar{n}^{a} \bar{n}_{c} \bar{\nabla}_{d}\bar{\nabla}_{a}L - 2 \bar{\nabla}_{d}\bar{\nabla}_{c}L.
\eeqn

\section{Decomposition of the linearized Einstein tensor}\label{app:LEinsteinTensor}
The linearized Einstein tensor is expressed as
\beqn
    G^{(1)}_{ab}=\frac{1}{2} \Bigl(\bar{\nabla}_{c}\bar{\nabla}_{a}h_{b}{}^{c} + \bar{\nabla}_{c}\bar{\nabla}_{b}h_{a}{}^{c} - h_{ab}{}^{(4)}{}\bar{R} -  \bar{\nabla}_{b}\bar{\nabla}_{a}h^{c}{}_{c} -  \bar{\nabla}_{c}\bar{\nabla}^{c}h_{ab} + \bar{g}_{ab} (h^{cd}{}^{(4)}\bar{R}_{cd} -  \bar{\nabla}_{d}\bar{\nabla}_{c}h^{cd} + \bar{\nabla}_{d}\bar{\nabla}^{d}h^{c}{}_{c})\Bigr).
    \label{eq:LEinsteinTensor}
\eeqn
Under the gauge condition (\ref{eq:gauge_condition}), substituting Eq. (\ref{eq:SVT}) into Eq. (\ref{eq:LEinsteinTensor}) and using the transverse condition and Eqs. (\ref{eq:GradNormal1})-(\ref{eq:GaussCodazziRicci}) to reduce the derivative acted on the perturbation as much as possible, the linearized Einstein tensor can be written as
\beqn
    G^{(1)}_{ab}\barn^a\barn^b
    &=&\Biggl[
        -\bargamma^{ab}\barnabla_a \barnabla_b C
    \Biggr]
    +\Biggl[ 
        -  \frac{1}{2} \bar{K}^{bc} \bar{n}^{a} \bar{\nabla}_{a}E^{\T}_{bc} 
        -  \bar{K}^{ab} \bar{\nabla}_{b}B^{\V}_{a}
    \Biggr]
    +\Biggl[
        - \frac{1}{2}{^{(3)}}\bar{R} A 
        -  \frac{1}{2}{^{(3)}}\bar{R} C 
        \nn \\ &&
        -  \frac{1}{2}{^{(3)}}\bar{R}^{ab} E^{\T}_{ab} 
        + \bar{a}^{a} \bar{K} B^{\V}_{a} 
        + \bar{a}^{a} \bar{K}_{ab} B^{\V b} 
        + 2 B^{\V}_{a} \bar{\nabla}^{a}\bar{K} 
        - 2 B^{\V}_{a} \bar{\nabla}_{b}\bar{K}^{ab}
    \Biggr],\label{eq:LEinsteinTensornn}
\eeqn
\beqn
    G^{(1)}_{bc}\barn^b \bargamma^c_{\ a}
    &=&\Biggl[
        -\bargamma_a^{\ b}\barn^c\barnabla_c \barnabla_b C 
        + \frac{1}{2}\bargamma^{bc}\barnabla_c \barnabla_b B^\V_a
    \Biggr]\nonumber \\
    && + 2 \Biggl[ 
        -  \bar{K} \bar{\nabla}_{a}A 
        -  \bar{K} \bar{\nabla}_{a}C 
        -  \bar{K}^{bc} \bar{\nabla}_{a}E^{\T}_{bc} 
        -  \bar{\nabla}_{a}B^{\V}_{b} 
        -  \bar{K} \bar{n}_{a} \bar{n}^{b} \bar{\nabla}_{b}A 
        -   \bar{K} \bar{n}_{a} \bar{n}^{b} \bar{\nabla}_{b}C 
        \nn \\ &&
        -   \bar{K}^{cd} \bar{n}_{a} \bar{n}^{b} \bar{\nabla}_{b}E^{\T}_{cd} 
        +  \bar{a}^{b} \bar{\nabla}_{b}B^{\V}_{a} 
        +  \bar{K} \bar{n}^{b} \bar{\nabla}_{b}B^{\V}_{a}
        +  \bar{K}_{ab} \bar{\nabla}^{b}A 
        +  \bar{K}_{ab} \bar{\nabla}^{b}C  
        + 2\bar{K}^{bc} \bar{\nabla}_{c}E^{\T}_{ab} 
        \nn \\ &&
        -  \bar{a}^{b} \bar{n}^{c} \bar{\nabla}_{c}E^{\T}_{ab} 
        -  2\bar{a}^{b} \bar{n}_{a} \bar{n}^{c} \bar{\nabla}_{c}B^{\V}_{b} 
        -  2\bar{K}^{bc} \bar{n}_{a} \bar{\nabla}_{c}B^{\V}_{b}  
    \Biggr]\nonumber \\
    && + 2 \Biggl[
        -  \bar{a}^{b} \bar{K} E^{\T}_{ab}
        -   \bar{a}^{b} \bar{K}_{b}^{c} E^{\T}_{ac} 
        -   \bar{a}^{b} \bar{K}_{a}^{c} E^{\T}_{bc} 
        +  \bar{a}_{a} \bar{K}^{bc} E^{\T}_{bc} 
        +  \bar{a}^{b} \bar{a}^{c} \bar{n}_{a} E^{\T}_{bc} 
        -   \bar{a}_{b} \bar{a}^{b} B^{\V}_{a}
        \nn \\ && 
        +  \bar{K}^2 B^{\V}_{a} 
        +  \bar{K}_{bc} \bar{K}^{bc} B^{\V}_{a} 
        + {^{(3)}}\bar{R} B^{\V}_{a} 
        + 2\bar{a}_{a} \bar{a}^{b} B^{\V}_{b} 
        -  2\bar{K} \bar{K}_{a}{}^{b} B^{\V}_{b} 
        -   \bar{K}_{ac} \bar{K}^{bc} B^{\V}_{b} 
        \nn \\ &&
        -  {^{(3)}}\bar{R}_{a}{}^{b} B^{\V}_{b} 
        -   B^{\V}_{b} \bar{\nabla}_{a}\bar{a}^{b} 
        +  E^{\T}_{bc} \bar{\nabla}_{a}\bar{K}^{bc} 
        + 2 B^{\V}_{b} \bar{\nabla}^{b}\bar{a}_{a} 
        -   B^{\V}_{a} \bar{\nabla}_{b}\bar{a}^{b} 
        -   \bar{n}_{a} \bar{n}^{b} B^{\V}_{c} \bar{\nabla}_{b}\bar{a}^{c} 
        \nn \\ &&
        + 2\bar{n}^{b} B^{\V}_{a} \bar{\nabla}_{b}\bar{K} 
        -  2\bar{n}^{b} B^{\V}_{c} \bar{\nabla}_{b}\bar{K}_{a}{}^{c} 
        +  \bar{n}_{a} \bar{n}^{b} E^{\T}_{cd} \bar{\nabla}_{b}\bar{K}^{cd} 
        -   E^{\T}_{ab} \bar{\nabla}^{b}\bar{K} 
        \nn \\ &&
        -  2E^{\T}_{bc} \bar{\nabla}^{c}\bar{K}^{b}{}_{a} 
        + 2E^{\T}_{ab} \bar{\nabla}_{c}\bar{K}^{bc} 
        - \bar{n}_{a} B^{\V}_{b} \bar{\nabla}_{c}\bar{K}^{bc}
    \Biggr],\label{eq:LinearizedEinsteintensorngamma}
\eeqn
\beqn
    G^{(1)}_{ab}\bargamma^{ab} 
    &=&\Biggl[
        -\bargamma^{ab}\barnabla_a\barnabla_b A 
        + \bargamma^{ab}\barnabla_a\barnabla_b C 
        -3\barn^a \barn^b \barnabla_b \barnabla_a C 
    \Biggr]\nonumber \\
    &&+\Biggl[ 
        - 2 \bar{a}^{a} \bar{\nabla}_{a}A 
        - 2 \bar{K} \bar{n}^{a} \bar{\nabla}_{a}A 
        - 2 \bar{a}^{a} \bar{\nabla}_{a}C 
        - 2 \bar{K} \bar{n}^{a} \bar{\nabla}_{a}C 
        -  \frac{3}{2} \bar{K}^{bc} \bar{n}^{a} \bar{\nabla}_{a}E^{\T}{}_{bc} 
        - 2 \bar{a}^{a} \bar{n}^{b} \bar{\nabla}_{b}B^{\V}_{a} 
        - 3 \bar{K}_{ab} \bar{\nabla}^{b}B^{\V}{}^{a}
    \Biggr]\nonumber \\
    &&+\Biggl[
        - \frac{1}{2} \bar{K}^2 A 
        -  \frac{3}{2} \bar{K}_{ab} \bar{K}^{ab} A 
        -  \frac{1}{2} \bar{K}^2 C 
        -  \frac{3}{2} \bar{K}_{ab} \bar{K}^{ab} C 
        - 3 \bar{a}^{a} \bar{a}^{b} E^{\T}_{ab} 
        + \bar{K} \bar{K}^{ab} E^{\T}_{ab}
        \nn \\ && 
        + \frac{3}{2}{^{(3)}}\bar{R}^{ab} E^{\T}_{ab} 
        -  \bar{a}^{a} \bar{K} B^{\V}_{a} 
        - 3 \bar{a}_{a} \bar{K}^{ab} B^{\V}_{b} 
        - 2 \bar{n}^{a} B^{\V}_{b} \bar{\nabla}_{a}\bar{a}^{b} 
        - 2 \bar{n}^{a} A \bar{\nabla}_{a}\bar{K} 
        - 2 \bar{n}^{a} C \bar{\nabla}_{a}\bar{K} 
        \nn \\ &&
        - 2 B^{\V}_{a} \bar{\nabla}^{a}\bar{K} 
        + \bar{n}^{a} E^{\T}_{bc} \bar{\nabla}_{a}\bar{K}^{bc} 
        - 3 E^{\T}_{ab} \bar{\nabla}^{b}\bar{a}^{a} 
    \Biggr],\label{eq:LinearizedEinsteintensortrace}
\eeqn
\beqn
    &&G^{(1)}_{cd}\left(\bargamma^{c}_{\ a}\bargamma^{d}_{\ b}-\frac{1}{3}\bargamma^{cd}\bargamma_{ab}\right)\nonumber \\
    &=&\Biggl[
        -\frac{1}{2}\barg^{cd}\barnabla_c \barnabla_d E^{\T}_{ab}+\frac{1}{2}\Bigl(\bargamma_a^{\ c}\bargamma_b^{\ d}-\frac{1}{3}\bargamma_{ab}\bargamma^{cd}\Bigr)\barnabla_c \barnabla_d (A-C)+\barn_{(a|} \barn^c \barn^d \barnabla_c\barnabla_d B^\V_{|b)} + \barn^c \barnabla_c \barnabla_{(a}B^\V_{b)}
    \Biggr]\nonumber \\
    &&+\Biggl[ 
        + \frac{1}{2} \bar{a}_{b} \bar{\nabla}_{a}A 
        + \frac{1}{2} \bar{a}_{b} \bar{\nabla}_{a}C 
        + \frac{1}{2} \bar{a}^{c} \bar{\nabla}_{a}E^{\T}_{bc} 
        + \frac{1}{2} \bar{K} \bar{\nabla}_{a}B^{\V}_{b} 
        -  \frac{1}{2} \bar{a}^{c} \bar{n}_{b} \bar{\nabla}_{a}B^{\V}_{c} 
        \nn \\ &&
        + \frac{1}{2} \bar{K}_{bc} \bar{\nabla}_{a}B^{\V}{}^{c} 
        + \frac{1}{2} \bar{a}_{a} \bar{\nabla}_{b}A 
        + \frac{1}{2} \bar{a}_{a} \bar{\nabla}_{b}C 
        + \frac{1}{2} \bar{a}^{c} \bar{\nabla}_{b}E^{\T}_{ac} 
        + \frac{1}{2} \bar{K} \bar{\nabla}_{b}B^{\V}_{a} 
        -  \frac{1}{2} \bar{a}^{c} \bar{n}_{a} \bar{\nabla}_{b}B^{\V}_{c} 
        \nn \\ &&
        + \frac{1}{2} \bar{K}_{ac} \bar{\nabla}_{b}B^{\V}{}^{c} 
        -  \frac{1}{3} \bar{a}^{c} \bar{\gamma}_{ab} \bar{\nabla}_{c}A 
        + \bar{K}_{ab} \bar{n}^{c} \bar{\nabla}_{c}A 
        -  \frac{1}{3} \bar{K} \bar{\gamma}_{ab} \bar{n}^{c} \bar{\nabla}_{c}A 
        + \frac{1}{2} \bar{a}_{b} \bar{n}_{a} \bar{n}^{c} \bar{\nabla}_{c}A 
        \nn \\ &&
        + \frac{1}{2} \bar{a}_{a} \bar{n}_{b} \bar{n}^{c} \bar{\nabla}_{c}A 
        -  \frac{1}{3} \bar{a}^{c} \bar{\gamma}_{ab} \bar{\nabla}_{c}C 
        + \bar{K}_{ab} \bar{n}^{c} \bar{\nabla}_{c}C 
        -  \frac{1}{3} \bar{K} \bar{\gamma}_{ab} \bar{n}^{c} \bar{\nabla}_{c}C 
        + \frac{1}{2} \bar{a}_{b} \bar{n}_{a} \bar{n}^{c} \bar{\nabla}_{c}C 
        \nn \\ &&
        + \frac{1}{2} \bar{a}_{a} \bar{n}_{b} \bar{n}^{c} \bar{\nabla}_{c}C 
        + \bar{a}_{b} \bar{n}^{c} \bar{\nabla}_{c}B^{\V}_{a} 
        + \frac{1}{2} \bar{K} \bar{n}_{b} \bar{n}^{c} \bar{\nabla}_{c}B^{\V}_{a} 
        + \bar{a}_{a} \bar{n}^{c} \bar{\nabla}_{c}B^{\V}_{b} 
        \nn \\ &&
        + \frac{1}{2} \bar{K} \bar{n}_{a} \bar{n}^{c} \bar{\nabla}_{c}B^{\V}_{b} 
        -  \frac{1}{2} \bar{K}_{bc} \bar{\nabla}^{c}B^{\V}_{a} 
        -  \frac{1}{2} \bar{K}_{ac} \bar{\nabla}^{c}B^{\V}_{b} 
        + \bar{K}^{cd} \bar{n}_{b} \bar{\nabla}_{d}E^{\T}_{ac} 
        -  \frac{1}{2} \bar{a}^{c} \bar{n}_{b} \bar{n}^{d} \bar{\nabla}_{d}E^{\T}_{ac} 
        \nn \\ &&
        + \bar{K}^{cd} \bar{n}_{a} \bar{\nabla}_{d}E^{\T}_{bc} 
        -  \frac{1}{2} \bar{a}^{c} \bar{n}_{a} \bar{n}^{d} \bar{\nabla}_{d}E^{\T}_{bc} 
        -  \frac{1}{3} \bar{a}^{c} \bar{\gamma}_{ab} \bar{n}^{d} \bar{\nabla}_{d}B^{\V}_{c} 
        -  \bar{a}^{c} \bar{n}_{a} \bar{n}_{b} \bar{n}^{d} \bar{\nabla}_{d}B^{\V}_{c} 
    \Biggr]\nonumber \\
    &&+\Biggl[
        \bar{K} \bar{K}_{ab} A 
        -  \frac{1}{3} \bar{K}^2 \bar{\gamma}_{ab} A 
        -  \bar{a}^{c} \bar{K}_{bc} \bar{n}_{a} A 
        -  \bar{a}^{c} \bar{K}_{ac} \bar{n}_{b} A 
        + \bar{K} \bar{K}_{ab} C 
        -  \frac{1}{3} \bar{K}^2 \bar{\gamma}_{ab} C 
        \nn \\ &&
        -  \bar{a}^{c} \bar{K}_{bc} \bar{n}_{a} C 
        -  \bar{a}^{c} \bar{K}_{ac} \bar{n}_{b} C 
        -  \frac{1}{2} \bar{K}^2 E^{\T}_{ab} 
        -  \frac{1}{2} \bar{K}_{cd} \bar{K}^{cd} E^{\T}_{ab} 
        -  \frac{1}{2}{^{(3)}}\bar{R} E^{\T}_{ab} 
        -  \frac{1}{2} \bar{a}_{b} \bar{a}^{c} E^{\T}_{ac} 
        \nn \\ &&
        + \frac{1}{2} \bar{K} \bar{K}_{b}{}^{c} E^{\T}_{ac} 
        -  \frac{1}{2} \bar{a}^{c} \bar{K} \bar{n}_{b} E^{\T}_{ac} 
        + \frac{1}{2}{^{(3)}}\bar{R}_{b}{}^{c} E^{\T}_{ac} 
        -  \frac{1}{2} \bar{a}_{a} \bar{a}^{c} E^{\T}_{bc} 
        + \frac{1}{2} \bar{K} \bar{K}_{a}{}^{c} E^{\T}_{bc} 
        -  \frac{1}{2} \bar{a}^{c} \bar{K} \bar{n}_{a} E^{\T}_{bc} 
        \nn \\ &&
        + \frac{1}{2}{^{(3)}}\bar{R}_{a}{}^{c} E^{\T}_{bc} 
        + \bar{K}_{a}{}^{c} \bar{K}_{b}{}^{d} E^{\T}_{cd} 
        -  \frac{1}{3} \bar{K} \bar{K}^{cd} \bar{\gamma}_{ab} E^{\T}_{cd} 
        -  \frac{1}{2} \bar{a}^{c} \bar{K}_{b}{}^{d} \bar{n}_{a} E^{\T}_{cd} 
        \nn \\ &&
        -  \frac{1}{2} \bar{a}^{c} \bar{K}_{a}{}^{d} \bar{n}_{b} E^{\T}_{cd} 
        + \bar{a}^{c} \bar{a}^{d} \bar{n}_{a} \bar{n}_{b} E^{\T}_{cd} 
        - {^{(3)}}\bar{R}_{acbd} E^{\T}{}^{cd} 
        -  \bar{K}_{c}{}^{e} \bar{K}^{cd} \bar{n}_{a} \bar{n}_{b} E^{\T}_{de} 
        \nn \\ &&
        + \frac{1}{2} \bar{a}_{b} \bar{K} B^{\V}_{a} 
        -  \frac{1}{2} \bar{a}^{c} \bar{K}_{bc} B^{\V}_{a} 
        -  \frac{1}{2} \bar{a}_{c} \bar{a}^{c} \bar{n}_{b} B^{\V}_{a} 
        + \frac{1}{2} \bar{a}_{a} \bar{K} B^{\V}_{b} 
        -  \frac{1}{2} \bar{a}^{c} \bar{K}_{ac} B^{\V}_{b} 
        -  \frac{1}{2} \bar{a}_{c} \bar{a}^{c} \bar{n}_{a} B^{\V}_{b} 
        \nn \\ &&
        + \bar{a}^{c} \bar{K}_{ab} B^{\V}_{c} 
        -  \frac{2}{3} \bar{a}^{c} \bar{K} \bar{\gamma}_{ab} B^{\V}_{c} 
        -  \frac{1}{2} \bar{a}_{b} \bar{a}^{c} \bar{n}_{a} B^{\V}_{c} 
        -  \frac{1}{2} \bar{a}_{a} \bar{a}^{c} \bar{n}_{b} B^{\V}_{c} 
        -  \frac{1}{2} \bar{K} \bar{K}_{bc} \bar{n}_{a} B^{\V}{}^{c} 
        \nn \\ &&
        -  \frac{1}{2} \bar{K}_{b}{}^{d} \bar{K}_{cd} \bar{n}_{a} B^{\V}{}^{c} 
        -  \frac{1}{2} \bar{K} \bar{K}_{ac} \bar{n}_{b} B^{\V}{}^{c} 
        -  \frac{1}{2} \bar{K}_{a}{}^{d} \bar{K}_{cd} \bar{n}_{b} B^{\V}{}^{c} 
        \nn \\ &&
        + \bar{a}^{c} \bar{K}_{cd} \bar{n}_{a} \bar{n}_{b} B^{\V}{}^{d} 
        + \frac{1}{2} E^{\T}_{bc} \bar{\nabla}_{a}\bar{a}^{c} 
        + \frac{1}{2} E^{\T}_{ac} \bar{\nabla}_{b}\bar{a}^{c} 
        + \frac{1}{2} \bar{n}^{c} B^{\V}_{b} \bar{\nabla}_{c}\bar{a}_{a} 
        \nn \\ &&
        + \frac{1}{2} \bar{n}^{c} B^{\V}_{a} \bar{\nabla}_{c}\bar{a}_{b} 
        + E^{\T}_{ab} \bar{\nabla}_{c}\bar{a}^{c} 
        -  \frac{1}{3} \bar{\gamma}_{ab} \bar{n}^{c} B^{\V}{}^{d} \bar{\nabla}_{c}\bar{a}_{d} 
        -  \frac{1}{3} \bar{\gamma}_{ab} \bar{n}^{c} A \bar{\nabla}_{c}\bar{K} 
        \nn \\ &&
        -  \frac{1}{3} \bar{\gamma}_{ab} \bar{n}^{c} C \bar{\nabla}_{c}\bar{K} 
        -  \bar{n}^{c} E^{\T}_{ab} \bar{\nabla}_{c}\bar{K} 
        -  \frac{1}{3} \bar{\gamma}_{ab} B^{\V}{}^{c} \bar{\nabla}_{c}\bar{K} 
        + \bar{n}^{c} A \bar{\nabla}_{c}\bar{K}_{ab} 
        + \bar{n}^{c} C \bar{\nabla}_{c}\bar{K}_{ab} 
        \nn \\ &&
        + B^{\V}{}^{c} \bar{\nabla}_{c}\bar{K}_{ab} 
        + \frac{1}{2} \bar{n}^{c} E^{\T}_{b}{}^{d} \bar{\nabla}_{c}\bar{K}_{ad} 
        -  \frac{1}{2} \bar{n}_{b} \bar{n}^{c} B^{\V}{}^{d} \bar{\nabla}_{c}\bar{K}_{ad} 
        + \frac{1}{2} \bar{n}^{c} E^{\T}_{a}{}^{d} \bar{\nabla}_{c}\bar{K}_{bd} 
        \nn \\ &&
        -  \frac{1}{2} \bar{n}_{a} \bar{n}^{c} B^{\V}{}^{d} \bar{\nabla}_{c}\bar{K}_{bd} 
        -  \frac{1}{3} \bar{\gamma}_{ab} \bar{n}^{c} E^{\T}{}^{de} \bar{\nabla}_{c}\bar{K}_{de} 
        -  \frac{1}{2} E^{\T}_{bc} \bar{\nabla}^{c}\bar{a}_{a} 
        -  \frac{1}{2} E^{\T}_{ac} \bar{\nabla}^{c}\bar{a}_{b} 
        \nn \\ &&
        + \frac{1}{2} \bar{n}_{b} E^{\T}_{a}{}^{c} \bar{\nabla}_{d}\bar{K}_{c}{}^{d} 
        + \frac{1}{2} \bar{n}_{a} E^{\T}_{b}{}^{c} \bar{\nabla}_{d}\bar{K}_{c}{}^{d}
    \Biggr].\label{eq:LEinsteinTensortraceless}
\eeqn

\bibliographystyle{JHEPmod.bst}
\bibliography{refs.bib}

\end{document}